\documentclass{ws_tasi}

\def\bit{\begin{itemize}}                                                      
\def\eit{\end{itemize}}
\newcommand{\met}{\mbox{${E\!\!\!/}_T$}}
\def\bqa{\begin{eqnarray}}
\def\eqa{\end{eqnarray}}

\makeatletter

\newcommand\@pnumwidth{1.55em}
\newcommand\@tocrmarg{2.55em}
\newcommand\@dotsep{4.5}

\newcommand\tableofcontents{%
{\global
\@topnum\z@
\@afterindentfalse
\if@twocolumn
\@restonecoltrue
\onecolumn
\else
\@restonecolfalse
\fi
\vspace*{24pt}
\noindent
{\small\bf Contents}\par
\vskip1em
\nobreak}
{\small
\@starttoc{toc}%
}\if@restonecol
\twocolumn
\fi}

\newcommand*\l@section[2]{%
   \ifnum \c@tocdepth >\z@
     \addpenalty\@secpenalty
     \addvspace{1.0em \@plus\p@}%
     \setlength\@tempdima{1.5em}%
     \begingroup
       \parindent \z@ \rightskip \@pnumwidth
       \parfillskip -\@pnumwidth
       \leavevmode \bfseries
       \advance\leftskip\@tempdima
       \hskip -\leftskip
       #1\nobreak\hfil \nobreak\hb@xt@\@pnumwidth{\hss #2}\par
     \endgroup
   \fi}

   \newcommand*\l@subsection{\@dottedtocline{2}{1.5em}{2.3em}}

\makeatother

\begin{document}

\title{TASI LECTURES ON PRECISION ELECTROWEAK PHYSICS}

\author{KONSTANTIN MATCHEV
}

\address{Department of Physics, \\
         University of Florida, \\
         Gainesville, FL 32611, USA\\
                and\\
Institute for High Energy Phenomenology,\\
Newman Laboratory for Elementary Particle Physics, \\
Cornell University, Ithaca, NY 14853, USA\\
E-mail: matchev@phys.ufl.edu}

\maketitle

\abstracts{
\vskip -3.5in
\rightline{hep-ph/0402031}
\rightline{UFIFT-HEP-03-11}
\rightline{CLNS 04/1862}
\rightline{January 2004}
\vskip 3.5in
These notes are a written version of a set of lectures given at 
TASI-02 on the topic of precision electroweak physics.}

\newpage
\tableofcontents

\newpage

\section{Preliminary Remarks}

Precision experiments have been crucial in both validating the 
Standard Model (SM) of particle physics and in providing directions
in searching for new physics. The precision program, which started 
with the discovery of the weak neutral currents in 1973,
has been extremely successful: it confirmed the gauge principle
in the Standard Model, established the gauge groups and 
representations and tested the one-loop structure of the SM,
validating the basic principles of renormalization,
which in turn allowed for a prediction of the top quark 
and Higgs boson masses. In relation to new physics models, 
precision measurements have severely constrained new physics
at the TeV scale and provided a hint of a possible (supersymmetric)
gauge coupling unification at high energies.

\subsection{What is it all about?}

On the one hand, the term {\em Precision Electroweak Physics} (PEW) refers to 
quantities which are very well measured. How well? Let us say at the 
\% level or better. However, not every well-measured 
quantity is of interest here. For example, 
the amount of cold dark matter in the Universe
$\Omega_{CDM}$, Newton's constant $G_N$ and the strong coupling constant 
$\alpha_s$ are all very well known, yet they 
do not belong to the {\em electroweak} sector of the Standard Model:
\begin{eqnarray}
SU(3) &\times& SU(2)_W \times U(1)_Y \nonumber \\
{\rm QCD}   && {\rm ~Electroweak}  \nonumber
\end{eqnarray}
The observables which are typically included in the
precision electroweak data set are conveniently summarized 
in Tables~\ref{table_zpole} and~\ref{table_nonzpole}.

Our goal in these lectures will be the following:
\begin{enumerate}
\item we shall learn the meaning of some of the quantities in Tables~\ref{table_zpole}
and \ref{table_nonzpole};
\item we shall find out how these quantities are measured by experiment;
\item we shall discuss the prospects for improving these measurements in the near future;
\item we shall find out their implications about the validity of the Standard Model;
\item we shall learn how they can be used to predict or constrain as yet 
undiscovered physics -- for example, the mass of the Higgs boson, 
or the existence/lack of new physics beyond the Standard Model near the TeV scale.
\end{enumerate}

\begin{table}[tbp]
\tbl{Principal $Z$-pole observables, their experimental values, 
theoretical predictions using the SM parameters from the global best fit
as of 1/03, and pull (difference from the prediction divided by the theoretical uncertainty).
$\Gamma({\rm had})$, $\Gamma({\rm inv})$, and $\Gamma({\ell^+\ell^-})$ are not independent,
but are included for completeness. From Ref.~\protect\refcite{Langacker:2003tv}.
\label{table_zpole}}
{
\begin{tabular}{|l|c|c|c|r|}
\hline Quantity & Group(s) & Value & Standard Model & pull \\ 
\hline
$M_Z$ \hspace{14pt}      [GeV]&     LEP     &$ 91.1876 \pm 0.0021 $&$ 91.1874 \pm 0.0021 $&$ 0.1$ \\
$\Gamma_Z$ \hspace{17pt} [GeV]&     LEP     &$  2.4952 \pm 0.0023 $&$  2.4972 \pm 0.0011 $&$-0.9$ \\
{ $\Gamma({\rm had})$\hspace{8pt}[GeV]}
                              &     LEP     &$  1.7444 \pm 0.0020 $&$  1.7436 \pm 0.0011 $&  ---  \\
{ $\Gamma({\rm inv})$ [MeV]}
                              &     LEP     &$499.0    \pm 1.5    $&$501.74   \pm 0.15   $&  ---  \\
{ $\Gamma({\ell^+\ell^-})$ [MeV]}
                              &     LEP     &$ 83.984  \pm 0.086  $&$ 84.019  \pm 0.027  $&  ---  \\
$\sigma_{\rm had}$ \hspace{12pt}[nb]&LEP    &$ 41.541  \pm 0.037  $&$ 41.470  \pm 0.010  $&$ 1.9$ \\
$R_e$                         &     LEP     &$ 20.804  \pm 0.050  $&$ 20.753  \pm 0.012  $&$ 1.0$ \\
$R_\mu$                       &     LEP     &$ 20.785  \pm 0.033  $&$ 20.753  \pm 0.012  $&$ 1.0$ \\
$R_\tau$                      &     LEP     &$ 20.764  \pm 0.045  $&$ 20.799  \pm 0.012  $&$-0.8$ \\
\hline
$A_{FB} (e)$                  &     LEP     &$  0.0145 \pm 0.0025 $&$ 0.01639 \pm 0.00026$&$-0.8$ \\
$A_{FB} (\mu)$                &     LEP     &$  0.0169 \pm 0.0013 $&$                    $&$ 0.4$ \\
$A_{FB} (\tau)$               &     LEP     &$  0.0188 \pm 0.0017 $&$                    $&$ 1.4$ \\
$R_b$                         &  LEP + SLD  &$  0.21664\pm 0.00065$&$  0.21572\pm 0.00015$&$ 1.1$ \\
$R_c$                         &  LEP + SLD  &$  0.1718 \pm 0.0031 $&$  0.17231\pm 0.00006$&$-0.2$ \\
$R_{s,d}/R_{(d+u+s)}$         &     OPAL    &$  0.371  \pm 0.023  $&$  0.35918\pm 0.00004$&$ 0.5$ \\
$A_{FB} (b)$                  &     LEP     &$  0.0995 \pm 0.0017 $&$  0.1036 \pm 0.0008 $&$-2.4$ \\
$A_{FB} (c)$                  &     LEP     &$  0.0713 \pm 0.0036 $&$  0.0741 \pm 0.0007 $&$-0.8$ \\
$A_{FB} (s)$                  &DELPHI,OPAL  &$  0.0976 \pm 0.0114 $&$  0.1037 \pm 0.0008 $&$-0.5$ \\
$A_b$                         &     SLD     &$  0.922  \pm 0.020  $&$  0.93476\pm 0.00012$&$-0.6$ \\
$A_c$                         &     SLD     &$  0.670  \pm 0.026  $&$  0.6681 \pm 0.0005 $&$ 0.1$ \\
$A_s$                         &     SLD     &$  0.895  \pm 0.091  $&$  0.93571\pm 0.00010$&$-0.4$ \\
\hline
$A_{LR}$ (hadrons)            &     SLD     &$  0.15138\pm 0.00216$&$  0.1478 \pm 0.0012 $&$ 1.7$ \\
$A_{LR}$ (leptons)            &     SLD     &$  0.1544 \pm 0.0060 $&$                    $&$ 1.1$ \\
$A_\mu$                       &     SLD     &$  0.142  \pm 0.015  $&$                    $&$-0.4$ \\
$A_\tau$                      &     SLD     &$  0.136  \pm 0.015  $&$                    $&$-0.8$ \\
$A_e (Q_{LR})$                &     SLD     &$  0.162  \pm 0.043  $&$                    $&$ 0.3$ \\
$A_\tau ({\mathcal P}_\tau)$  &     LEP     &$  0.1439 \pm 0.0043 $&$                    $&$-0.9$ \\
$A_e ({\mathcal P}_\tau)$     &     LEP     &$  0.1498 \pm 0.0048 $&$                    $&$ 0.4$ \\
$Q_{FB}$                      &     LEP     &$  0.0403 \pm 0.0026 $&$  0.0424 \pm 0.0003 $&$-0.8$ \\
\hline
\end{tabular}
}
\end{table}

\subsection{Useful references for further reading}

For the current status of the precision electroweak observables, one should consult
the ``Electroweak model and constraints on New Physics'' review~\cite{Erler:ew}
in the Particle Data Book~\cite{Hagiwara:fs}. 
There are nice lecture summaries from previous 
schools~\cite{Pich:1994zt,Langacker:1995ch,Pierce:1997wu,%
Hewett:1997zp,Dawson:1998yi,Kim:2000fh,Drees:2001xw,Renton:2002wy}, 
as well as book collections of relevant review articles~\cite{Langacker:qb,Barklow:gc,epatZ}.
The long-term prospects for improvement in the precision measurements
have been discussed in several workshop reports, e.g. the Tevatron Run II Workshop 
\cite{Brock:1999ep} and Snowmass 2001~\cite{Langacker:2001ij,Baur:2001yp,Baur:2002gp}.
The website of the LEP Electroweak Working Group 
is another excellent source of information, with 
its continuously updated electroweak summary notes~\cite{unknown:2003ih}.

\begin{table}[tbp] 
\tbl{The recent status of non-$Z$-pole observables, as of 1/03. 
From Ref.~\protect\refcite{Langacker:2003tv}.}
{
\begin{tabular}{|l|c|c|c|r|}
\hline Quantity & Group(s) & Value & Standard Model & pull \\ 
\hline
$m_t$\hspace{8pt}[GeV]&Tevatron &$ 174.3    \pm 5.1               $&$ 174.4    \pm 4.4    $&$ 0.0$ \\
$M_W$ [GeV]    &      LEP       &$  80.447  \pm 0.042             $&$  80.391  \pm 0.018  $&$ 1.3$ \\
$M_W$ [GeV]    &  Tevatron,UA2  &$  80.454  \pm 0.059             $&$                     $&$ 1.1$ \\
\hline
$g_L^2$        &     NuTeV      &$   0.30005\pm 0.00137           $&$   0.30396\pm 0.00023$&$-2.9$ \\
$g_R^2$        &     NuTeV      &$   0.03076\pm 0.00110           $&$   0.03005\pm 0.00004$&$ 0.6$ \\
$R^\nu$        &     CCFR       &$   0.5820 \pm 0.0027 \pm 0.0031 $&$   0.5833 \pm 0.0004 $&$-0.3$ \\
$R^\nu$        &     CDHS       &$   0.3096 \pm 0.0033 \pm 0.0028 $&$   0.3092 \pm 0.0002 $&$ 0.1$ \\
$R^\nu$        &     CHARM      &$   0.3021 \pm 0.0031 \pm 0.0026 $&$                     $&$-1.7$ \\
$R^{\bar\nu}$  &     CDHS       &$   0.384  \pm 0.016  \pm 0.007  $&$   0.3862 \pm 0.0002 $&$-0.1$ \\
$R^{\bar\nu}$  &     CHARM      &$   0.403  \pm 0.014  \pm 0.007  $&$                     $&$ 1.0$ \\
$R^{\bar\nu}$  &     CDHS 1979  &$   0.365  \pm 0.015  \pm 0.007  $&$   0.3816 \pm 0.0002 $&$-1.0$ \\
\hline
$g_V^{\nu e}$  &     CHARM II   &$  -0.035  \pm 0.017             $&$  -0.0398 \pm 0.0003 $&  ---  \\
$g_V^{\nu e}$  &      all       &$  -0.041  \pm 0.015             $&$                     $&$-0.1$ \\
$g_A^{\nu e}$  &     CHARM II   &$  -0.503  \pm 0.017             $&$  -0.5065 \pm 0.0001 $&  ---  \\
$g_A^{\nu e}$  &      all       &$  -0.507  \pm 0.014             $&$                     $&$ 0.0$ \\
\hline
{} &{} &{} &{} &{}\\[-2.0ex]
$Q_W({\rm Cs})$&     Boulder    &$ -72.69   \pm 0.44              $&$ -73.10   \pm 0.04   $&$ 0.8$ \\
$Q_W({\rm Tl})$&Oxford/Seattle  &$-116.6    \pm 3.7               $&$-116.7    \pm 0.1    $&$ 0.0$ \\
\hline
{} &{} &{} &{} &{}\\[-2.0ex]
$10^3$ $\frac{\Gamma (b\rightarrow s\gamma)}{\Gamma_{SL}}$ 
               &BaBar/Belle/CLEO&$ 3.48^{+0.65}_{-0.54}           $&$     3.20 \pm 0.09   $&$ 0.5$ \\
\hline
{} &{} &{} &{} &{}\\[-2.0ex]
$\tau_\tau$ [fs] & direct/${\mathcal B}_e/ {\mathcal B}_\mu$ &$ 290.96 \pm 0.59 \pm 5.66 $&$ 291.90 \pm 1.81 $&$-0.4$ \\
$10^4$ $\Delta\alpha^{(3)}_{\rm had}$ & $e^+e^-$/$\tau$ decays &$ 56.53 \pm 0.83 \pm 0.64 $&$ 57.52 \pm 1.31 $&$-0.9$ \\
$10^9$ $(a_\mu - \frac{\alpha}{2\pi})$ & BNL/CERN &$ 4510.64 \pm 0.79 \pm 0.51 $&$ 4508.30 \pm 0.33 $&{ $2.5$} \\
\hline
\end{tabular}
\label{table_nonzpole}
}
\end{table}


\section{The Tools of the Trade}

As usual, we need to have a theory and experiments to test it. 
Before we go on to discuss the observables from
Tables~\ref{table_zpole} and~\ref{table_nonzpole} in the next few sections, 
we first need to introduce the necessary theoretical ingredients
and describe the relevant type of experiments. In Section~\ref{sec:SMtheory}
we review the electroweak Lagrangian and in Section~\ref{sec:inputs} 
we discuss its input parameters. Sections~\ref{sec:experiments}
provides a general review of some of the types of experiments 
which have contributed to 
the results in Tables~\ref{table_zpole} and~\ref{table_nonzpole}.

\subsection{Theory}
\label{sec:SMtheory}

Achieving the remarkable precision displayed in 
Tables~\ref{table_zpole} and~\ref{table_nonzpole} is only possible
because the Standard Model provides a well-defined theoretical framework 
for computing the different electroweak observables in terms of a few input
parameters and thus predicting different relations among them.

The SM electroweak Lagrangian for fermions is given by
\begin{eqnarray}
{\mathcal L}_F &=& \sum_i \bar{\psi}_i
\left(i\slash\!\!\!\partial -m_i-\frac{g_im_ih}{2M_W} \right) 
\psi_i \label{Higgs}\\
&-& e \sum_i Q^i \bar{\psi}_i \gamma^\mu \psi_i A_\mu ~~~ (QED) \label{QED}\\
&-& \frac{g}{2\sqrt{2}} \sum_i \bar{\psi}_i \gamma^\mu (1-\gamma^5) 
\left(T^+W^+_\mu + T^-W^-_\mu\right)\psi_i  ~~~(CC) \label{CC}\\
&-& \frac{g}{2\cos\theta_W} \sum_i \bar{\psi}_i \gamma^\mu (g_V^i-g_A^i\gamma^5) 
\psi_i Z_\mu ~~~(NC) \label{NC}
\end{eqnarray}
where the vector and axial vector couplings are
\begin{eqnarray}
g_V^i &=& T_3^i - 2Q^i\sin^2\theta_W,  \label{gV}\\
g_A^i &=& T_3^i, \label{gA}
\end{eqnarray}
$Q^i$ is the electric charge of fermion $i$, $T_3^i$ is its weak isospin,
$\theta_W$ is the Weinberg angle ($\tan\theta_W=g'/g$),
$g$ ($g'$) is the $SU(2)_W$ ($U(1)_Y$) gauge coupling constant,
and $e=g\sin\theta_W$ is the electromagnetic coupling constant.
The photon ($A_\mu$) and Z-boson ($Z_\mu$) fields are 
given in terms of the hypercharge gauge boson $B_\mu$ and 
neutral $W$-boson $W^3_\mu$ as
\begin{eqnarray}
A_\mu &=& B_\mu\cos\theta_W + W^3_\mu\sin\theta_W, \label{A}\\
Z_\mu &=&-B_\mu\sin\theta_W + W^3_\mu\cos\theta_W. \label{Z}
\end{eqnarray}
Often the neutral current interactions are equivalently written as
\begin{equation}
-\frac{g}{2\cos\theta_W} \sum_i \bar{\psi}_i \gamma^\mu 
\left(g_L^i(1-\gamma^5)+g_R^i(1+\gamma^5)\right) 
\psi_i Z_\mu ~~~(NC) \label{NCLR}
\end{equation}
with the redefinition
\begin{eqnarray}
g_V &=& g_L + g_R  \\
g_A &=& g_L - g_R
\end{eqnarray}

{\em \underline{\bf Exercise} Use (\ref{gV}), (\ref{gA}), (\ref{A}) and (\ref{Z})
to derive (\ref{Higgs}-\ref{NC}) from the $SU(2)_W\times U(1)_Y$ gauge theory 
lagrangian supplemented with the Yukawa terms for the fermions. 
Hint: you will need to know the Higgs vacuum expectation value $v$,
which is computed by minimizing the tree-level Higgs potential
\begin{equation}
V_H = -\frac{1}{2}\mu^2 h^2 + \frac{\lambda_H}{4}h^4
\label{HiggsPotential}
\end{equation}
as
\begin{equation}
v=\sqrt{\frac{\mu^2}{\lambda_H}}.
\end{equation}
}

\subsection{Fundamental parameters, input parameters and observables}
\label{sec:inputs}

{\em Fundamental} parameters are the parameters appearing
in the original gauge theory Lagrangian. In the case of the electroweak
Lagrangian, the fundamental parameters are the gauge couplings
$g'$ and $g$. The Higgs sector of the theory (\ref{HiggsPotential})
adds two more fundamental parameters: the Higgs mass 
parameter $\mu$ and self-coupling $\lambda_H$. 
The remaining fundamental parameters are the Yukawa couplings in the
Yukawa sector (\ref{Higgs}). Putting all together, the set of 
fundamental parameters is
\begin{equation}
\{g,g',\lambda_H,\mu^2,\lambda_i\}.
\label{fundamental}
\end{equation} 

Often, as was the case above in Eq.~(\ref{Higgs}-\ref{NC}),
it is convenient to rewrite the Lagrangian in terms of derived quantities,
i.e. {\em input} parameters. Through a simple redefinition we can trade 
one fundamental parameter for a combination of a new (input) parameter and
the remaining fundamental parameters, e.g.
\begin{eqnarray}
g' = g \tan\theta_W
\end{eqnarray}
eliminates $g'$ from the Lagrangian and replaces it with $\theta_W$.
Another example is
\begin{eqnarray}
v = \frac{2M_W}{g},
\end{eqnarray}
which eliminates the combination $v=\mu/\sqrt{\lambda_H}$ 
in favor of the $W$-boson mass $M_W$. 
The latter is directly accessible experimentally,
unlike either $\lambda_H$ or $\mu$.
A similar trick allows to go from Yukawa couplings to fermion masses:
\begin{eqnarray}
\lambda_i = \frac{gm_i}{\sqrt{2}M_W},
\end{eqnarray}
which is very convenient, as most fermion masses are well-known 
experimentally while in contrast the Yukawa couplings are
small (with the exception of the top Yukawa) and thus difficult to
measure. Of course, one should make sure the total number of parameters
stays the same for both the fundamental and input parameters.

For the precision electroweak tests of the SM the following
set of well-measured input parameters is typically used:
\begin{equation}
\{\alpha,G_F,M_Z,m_h,m_i\},
\label{inputs}
\end{equation} 
where $\alpha=e^2/4\pi$ is the fine structure constant,
$G_F$ is the Fermi constant, which is measured from the muon lifetime,
and $M_Z$ ($m_h$) is the $Z$-boson (Higgs boson) mass. 

{\em \underline{\bf Exercise} Find the relation between 
the input parameter set (\ref{inputs}) and the fundamental 
parameter set (\ref{fundamental}).}

Finally, {\em observables} are the quantities which are measured by
experiment. An example is the set of electroweak observables listed
in Tables~\ref{table_zpole} and \ref{table_nonzpole}. If the 
number of observables is larger than the number of input parameters,
we can test the model. 

\subsection{Experimental facilities}
\label{sec:experiments}

As can be seen from Tables~\ref{table_zpole} 
and \ref{table_nonzpole}, the electroweak observables have been
measured at a variety of experimental facilities. 
We shall now briefly describe each type.

\vspace*{0.5cm}
\centerline{\bf Lepton colliders}
\vspace*{0.5cm}

Lepton ($e^+e^-$) colliders have played a very important role
in the precision program. Lepton colliders have numerous advantages, 
perhaps best summarized in the following aphorism:
{\em ``At a lepton collider, every event is signal. 
At a hadron collider, every event is background.''} 
The main attractive features are:
\begin{itemize}
\item Fixed center of mass energy in each event. 
This provides an additional kinematic constraint, useful for 
eliminating the backgrounds as well as extracting properties 
of new particles, allowing e.g. a missing mass measurement. 
\item All of the available beam energy is used 
(minus beamstrahlung), i.e. we have an efficient use 
of particle acceleration.
\item Clean environment: easier identification of the underlying physics
in the event, easier heavy flavor tagging, less backgrounds.  
\item Polarizability of the colliding beams (SLC).
The polarization of the colliding beams can be used to 
reweight the contribution of the different diagrams involving 
$W$'s, $Z$'s and $\gamma$'s.
\end{itemize}

Naturally, lepton colliders also have certain disadvantages
in comparison to hadron colliders (see below), which makes
both types of facilities equally valuable and 
to a large extent complementary.

\begin{figure}[tb]
\centerline{\epsfxsize=4.1in\epsfbox{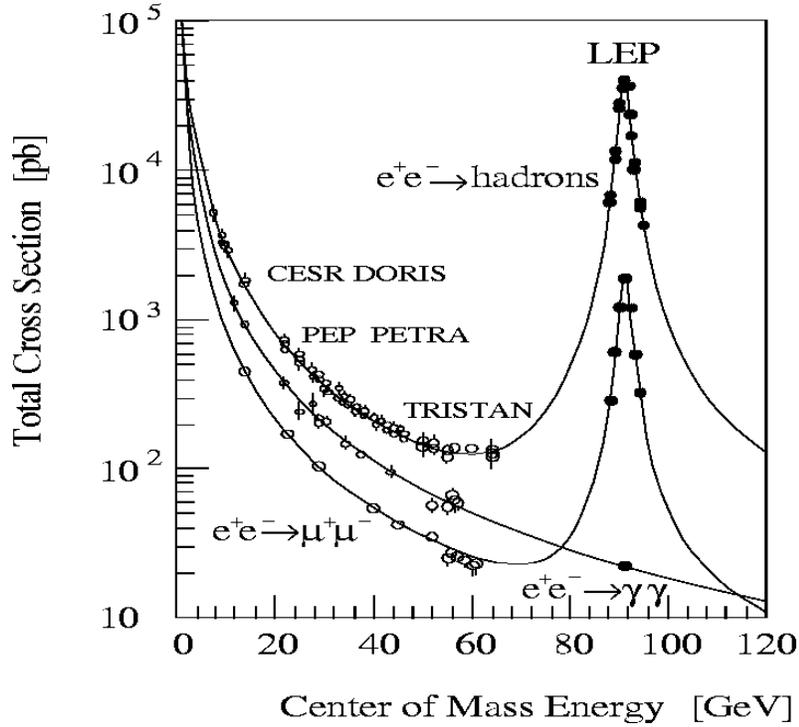}}   
\caption{Total $e^+e^-$ cross section from CESR/DORIS to LEP energies.}
\label{xsec}
\end{figure}
Figure~\ref{xsec} summarizes some of the history of $e^+e^-$ colliders. 
In reverse chronological order (including future facilities), they are:
\begin{itemize}
\item The Next Linear Collider (NLC) with an initial CM energy of 500 GeV
is now under active discussion as the next large international
particle physics facility. Its timescale is still uncertain, but
the NLC was ranked as the top priority among mid-term science facilities in the
Department of Energy's Office of Science 20-year science facility plan
\cite{doe}.
\item LEP-II at CERN (1996-2000). It had the highest CM energy 
so far among lepton colliders. LEP-II took data at several $\sqrt{s}$
before shutting down in November 2000 amidst much speculation and
heated discussions. There were four experiments (detectors) at
diametrically opposite sites: ALEPH, DELPHI, L3 and OPAL.
\item LEP-I at CERN (1989-1995), 
which took data at several energies around $\sqrt{s}=M_Z$.
\item SLC at SLAC (1989-1998). It reached energies up to 100 GeV
and also took data around the $Z$-pole. Unlike LEP, it had
an important advantage: the availability of beam polarization (80\%).
\item B-factories: PEP-II at SLAC (1999-present), $9\times3.1$ GeV;
KEKB at KEK (Japan) (1999-present) $8\times3.5$ GeV; 
CESR at Cornell (1979-present). {\em Exercise: Why are the B-factories asymmetric?}
\item Others ($\sqrt{s}<10$ GeV) (Novosibirsk, Bejing, Frascati).
\end{itemize}

\vspace*{0.5cm}
\centerline{\bf Hadron colliders}
\vspace*{0.5cm}

In turn, hadron colliders have their own advantages:
\begin{itemize}
\item Protons are much heavier than electrons, which leads to
a reduction in the radiation losses. As a result, for a given CM energy,
the ringsize of a hadron collider is smaller. Equivalently, 
for a given ringsize (and fixed magnetic field), a hadron collider
allows to reach higher CM energies.
\item Even though the CM energy of the colliding beams is large, 
the typical {\em parton} CM energy is much smaller, but 
usually still beyond the CM energy of LC competitors.
\end{itemize}

The main hadron collider facilities are the following:
\begin{itemize}
\item LHC (CERN) is currently under construction and 
is projected to begin operation in 2007-8 as a
$7\times7$ TeV $pp$ collider. 
There will be four experiments -- ATLAS, CMS, LHCB and ALICE.
The initial data taking rate will be 10 ${\rm fb}^{-1}$ per year 
at low luminosity and will subsequently increase 
to 100 ${\rm fb}^{-1}$ per year.
\item The Tevatron Run II at Fermilab is now operating as a
$0.98\times 0.98$ $p\bar{p}$ TeV collider. 
There are two experiments: CDF and D0, and the
hope nowadays is for 8 ${\rm fb}^{-1}$ per experiment before the LHC.
\item The Tevatron Run I (1987-1996): discovered the top quark.
It operated in a $0.9\times 0.9$ $p\bar{p}$ mode and delivered
110 ${\rm pb}^{-1}$ of data per experiment. 
\item $Sp\bar{p}S$ at CERN (1981-1990). It was a $300\times300$ GeV 
$p\bar{p}$ collider, and discovered the $W$ and $Z$ bosons.
\end{itemize}
It is interesting that the last three SM particles 
($W$, $Z$ and $t$) were discovered at hadron colliders
and chances are that the last one (the Higgs boson) will follow suit.

\vspace*{0.5cm}
\centerline{\bf Other facilities}
\vspace*{0.5cm}

In addition to the collider experiments mentioned above, there are 
numerous fixed target experiments, whose main advantage is 
the lower cost and large luminosity (because of the dense target).
Their main disadvantage is the low center of mass energy,
which scales with the beam energy $E_b$ only as  $E_{CM}\sim \sqrt{E_b}$.
Another class of experiments on muon dipole moments were done at
muon storage rings -- at CERN, and more recently, at BNL.

In conclusion of this section, we should comment on the experimental 
identification of heavy particles, which decay promptly and are
detected only through their decay products. For example, a
$W$-boson decays either to a pair of quarks, which later materialize into QCD jets,
or to a lepton and the corresponding neutrino. The $W$ branching fractions
are $B(W\rightarrow qq')\sim 2/3$, and
$B(W\rightarrow \ell\nu_\ell)\sim 1/9$ for $\ell=e,\mu,\tau$.
The jets (or the lepton and its neutrino) may have had a different
source, and the only indication that they may have come from a $W$ decay
is that their invariant mass is close to $M_W$.

Similarly, a $Z$-boson can decay to pairs of quarks, leptons or neutrinos,
with branching fractions $B(Z\rightarrow q\bar{q})\sim 0.7$, 
$B(Z\rightarrow \nu\bar{\nu})\sim 0.2$ and
$B(Z\rightarrow \ell^+\ell^-)\sim 0.1$ (summed over the three generations).
When both decay products are visible, their invariant mass is 
again required to be near $M_Z$.

A word of caution: $\tau$ is not always a lepton! 
A $\tau$-lepton can decay leptonically to $e$ or $\mu$ with a branching ratio
$0.18$ for each, or hadronically, with a branching ratio 0.64.
In the latter case it appears as a ``tau jet'', which is similar\footnote{Although
not quite -- a tau jet is more narrow and has lower track multiplicity.} 
to an ordinary QCD jet. To an experimentalist, a ``lepton'' is either an 
$e$ or a $\mu$, which may or may not have come from a tau decay. 
By $\tau$'s, experimentalists often mean ``tau jets''.

\section{Precision Measurements at the $Z$ Pole}

At the $Z$ pole the cross-section for $e^+e^-\rightarrow f\bar{f}$
is dominated by the $Z$ diagram. For $f\ne e$ we have
\begin{eqnarray}
\frac{d\sigma_Z^f}{d\Omega}=\frac{9}{4}
\frac{s\Gamma_{ee}\Gamma_{f\bar{f}}/M_Z^2}{(s-M_Z^2)^2+s^2\Gamma^2_Z/M_Z^2}
&\biggl[&(1+\cos^2\theta)(1-P_eA_e)  \nonumber \\
&+&2\cos\theta A_f(-P_e+A_e)\biggr]
\label{sigmaZ}
\end{eqnarray}
where $P_e$ is the polarization of the electron beam (relevant at SLC),
$s$ is the square of the CM energy, $\Gamma_Z$ is the total $Z$ 
width\footnote{The result (\ref{sigmaZ}) already incorporates the 
$s$-dependent width
$$
\Gamma_Z(s)=\frac{s}{M_Z^2}\, \Gamma_Z(s=M_Z^2),
$$
which in turn accounts for the effect of the so called ``non-photonic'',
or oblique, one-loop corrections, i.e. the corrections to the $Z$ propagator.
For details, see \protect\refcite{Montagna:1998sp}.},
$\Gamma_{ee}$ and $\Gamma_{f\bar{f}}$ are the partial 
widths for $Z\rightarrow ee$ and $Z\rightarrow f\bar{f}$,
correspondingly. (The partial widths are related to the $Z$ branching fractions
as $B(Z\rightarrow f\bar{f})=\Gamma_{f\bar{f}}/\Gamma_Z$.)
In (\ref{sigmaZ}) $\theta$ is the angle between the incident electron and the
outgoing fermion and $A_f$ is the left-right coupling constant asymmetry:
\begin{equation}
A_f = \frac{2g_V^fg_A^f}{(g_V^f)^2+(g_A^f)^2} = 
\frac{(g_L^f)^2-(g_R^f)^2}{(g_L^f)^2+(g_R^f)^2}
\end{equation}
where in the second equation we have used (\ref{gV}) and (\ref{gA}).
Notice that $A_f\le 1$. Since $g_V$ and $g_A$ only depend on the
quantum numbers of the particles, it follows that $A_f$ is the same for all
charged leptons, all up-type quarks and all down-type quarks.
For example, for charged leptons 
\begin{eqnarray}
g_V^\ell &=& -\frac{1}{2}-2(-1)\sin^2\theta_W \sim -0.50 + 0.462 = -0.038  \\
g_A^\ell &=& -\frac{1}{2} = -0.50
\end{eqnarray}
We see that because of an accidental cancellation $g_V^\ell \ll g_A^\ell$.
For the asymmetry we then get (at tree level)
\begin{equation}
A_\ell = \frac{2(-0.038)(-0.5)}{(-0.038)^2+(-0.5)^2} \sim 0.15
\end{equation}
(compare to the values measured in Table~\ref{table_zpole}). This small value of 
$A_\ell$ makes it particularly sensitive to electroweak vacuum polarization 
corrections (which have an impact on $\sin^2\theta_W$). In terms of
$\sin^2\theta_W$, we have
\begin{equation}
A_\ell = \frac{2(1-4\sin^2\theta_W)}{1+(1-4\sin^2\theta_W)^2}.
\end{equation}
Therefore small changes in $\sin^2\theta_W$ are amplified by a factor of 8
in the leptonic asymmetry $A_\ell$.

{\em {\bf Exercise:} Confirm roughly the numerical values for 
$A_b$ and $A_c$ in Table~\ref{table_zpole}.}

Let us now discuss the various measurements on the $Z$-pole. 
We shall need the following integrals for the forward and backward
regions:
\begin{eqnarray}
&F&:\quad 
\int_0      ^{\pi/2}(1+\cos^2\theta)\sin\theta d\theta = \frac{4}{3}; \qquad
\int_0      ^{\pi/2}2\cos\theta\sin\theta d\theta = 1; \label{intF}\\
&B&:\quad
\int_{\pi/2}^{\pi}(1+\cos^2\theta)\sin\theta d\theta = \frac{4}{3}; \qquad
\int_{\pi/2}^{\pi}2\cos\theta\sin\theta d\theta = -1. \label{intB}
\end{eqnarray}

\subsection{$Z$ resonance parameters}

\begin{figure}[tb]
\centerline{\epsfxsize=4.1in\epsfbox{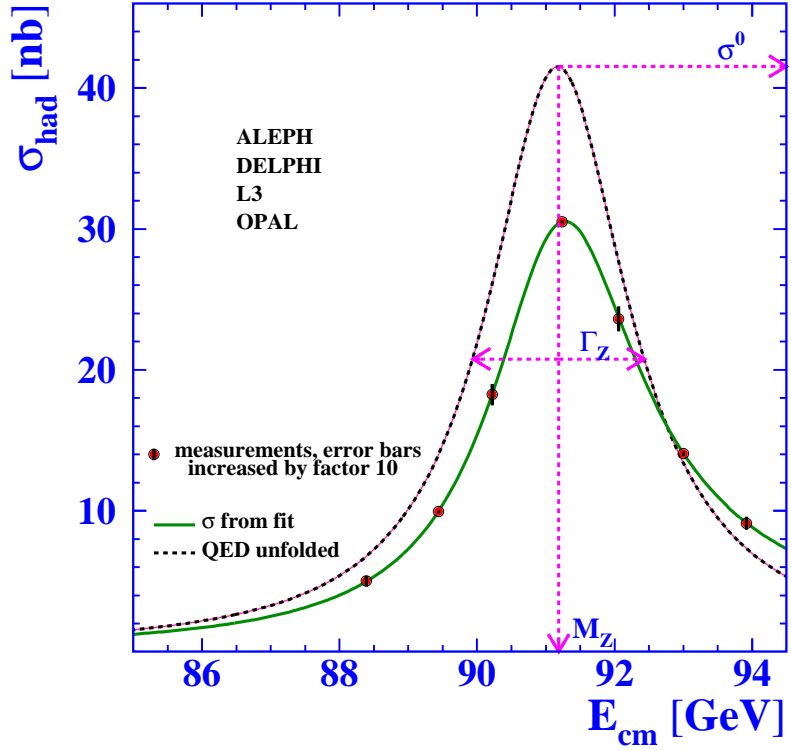}}   
\caption{The final hadronic cross-section averaged over the 
four experiments at LEP-I, as a function of the center-of-mass energy,
as measured (solid line) and QED deconvoluted (dashed line).}
\label{sig_had}
\end{figure}

Scanning the $Z$ peak and fitting to $\sigma_Z^f$ yields measurements of 
$M_Z$, $\Gamma_Z$ and the peak hadronic ($f=q$) total cross-section $\sigma_{had}$
(obtained for $s=M_Z^2$ by integrating (\ref{sigmaZ}) over the full solid angle):
\begin{equation}
\sigma_{had}=12\pi\frac{\Gamma_{ee}\Gamma_{had}}{M_Z^2\Gamma_Z^2}.
\end{equation}
Recall that the LEP beams are not polarized, so $P_e=0$. 
Figure~\ref{sig_had} shows the hadronic cross-section $\sigma_{had}$
measured by the four collaborations as a function of the center-of-mass 
energy (solid line). Also shown as a dashed line is the cross-section after
unfolding all effects due to photon radiation. The radiative corrections 
are large but very well known. At the peak the QED deconvoluted cross-section
is 36\% larger and the peak position is shifted by $\sim 100$ MeV.

\subsection{Branching ratios and partial widths}

Looking at various exclusive final states one can determine
the following ratios
\begin{eqnarray}
R_\ell &\equiv& \frac{\Gamma_{had}}{\Gamma_\ell} \sim \frac{0.7}{0.0333} \sim 21; \\ 
R_b    &\equiv& \frac{\Gamma_{bb}}{\Gamma_{had}}; \\ 
R_c    &\equiv& \frac{\Gamma_{cc}}{\Gamma_{had}}; 
\end{eqnarray}
with $\ell=\{e,\mu,\tau\}$. Basically this amounts to measuring the $Z$ branching fractions.

\begin{figure}[tb]
\centerline{\epsfxsize=4.1in\epsfbox{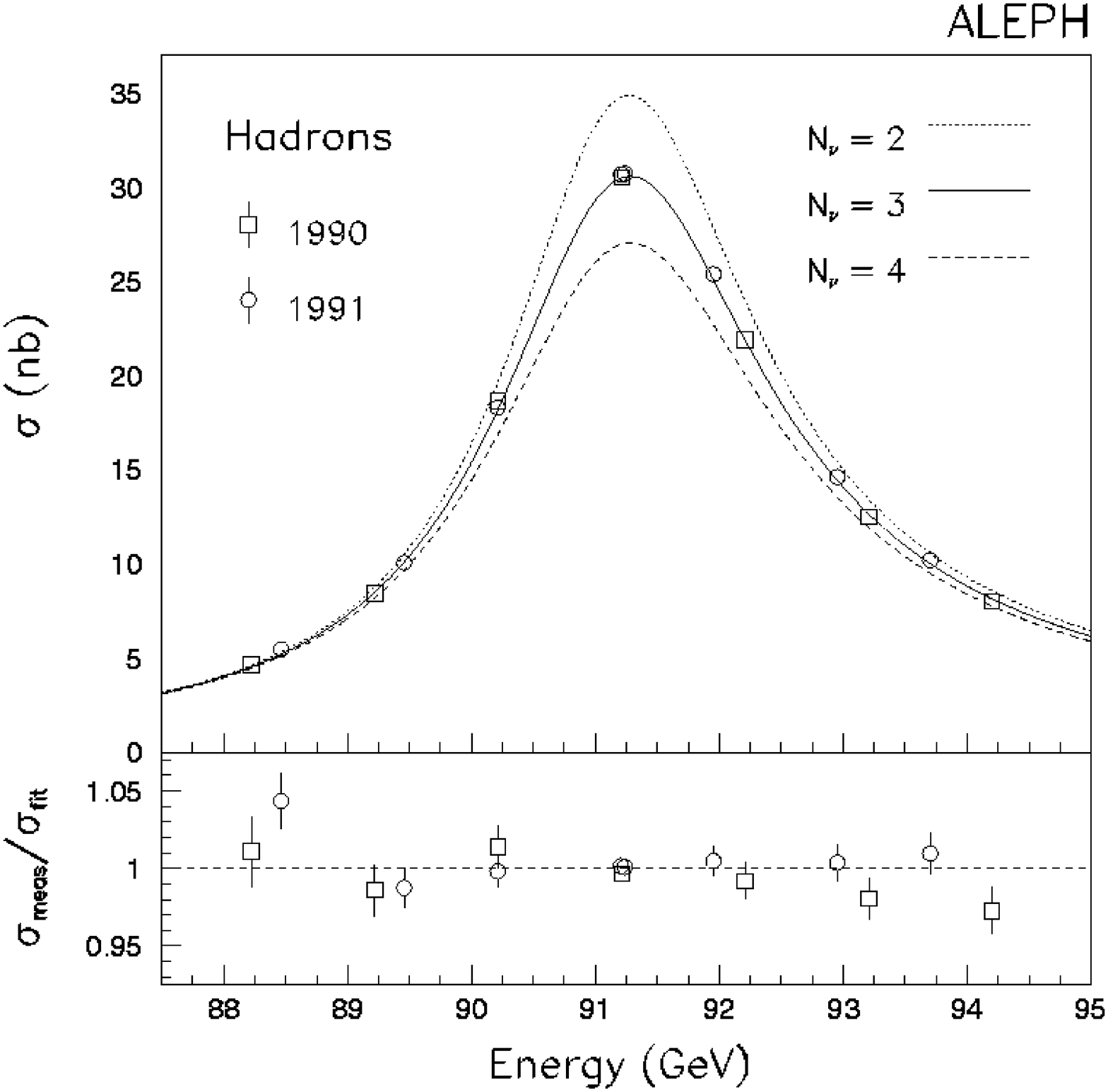}}   
\caption{The $Z$ lineshape in the hadronic channel along with the
SM prediction for two, three or four neutrinos species 
\protect\cite{Buskulic:1993gu}. The lower plot shows the ratio of the measured points
to the best-fit values. }
\label{3neutrinos}
\end{figure}

Having measured the total width from the peak scan and all visible partial widths,
one can determine the invisible partial width (for $Z\rightarrow \nu\bar{\nu}$):
\begin{equation}
\Gamma_{inv} = \Gamma_Z - \Gamma_{had} - \Gamma_{ee} - \Gamma_{\mu\mu} - \Gamma_{\tau\tau}
\end{equation}
and count the number of neutrino species $N_\nu$ coupling to the $Z$. 
Assuming that $\Gamma_{inv}=N_\nu \Gamma_\nu$, where $\Gamma_\nu$ is
the $Z$ partial width to pairs of a single neutrino species, we have
\begin{equation}
N_\nu = \frac{\Gamma_{inv}/\Gamma_\ell}{\left(\Gamma_\nu/\Gamma_\ell \right)_{SM}}.
\end{equation}
The ratio of $\Gamma_{inv}$ and the $Z$ partial width into charged leptons
is used in order to minimize the uncertainties due to the electroweak corrections 
which are common to both partial widths and cancel out in their ratio.
The $Z$ lineshape in the hadronic channel is shown in Figure~\ref{3neutrinos} 
along with the SM prediction for two, three or four neutrinos species.
The result confirms that there are three light neutrino flavors with
SM-like couplings to the $Z$.

\subsection{Unpolarized forward-backward asymmetry}

The unpolarized forward-backward asymmetry is determined by
a simple counting method in $e^+e^-\rightarrow f\bar{f}$ scattering
\begin{equation}
A_{FB}^f \equiv \frac{\sigma_F^f-\sigma_B^f}{\sigma_F^f+\sigma_B^f} 
= \frac{3}{4}A_eA_f,
\label{afb}
\end{equation}
where $\sigma_F^f$ ($\sigma_B^f$) is the total cross-section
for forward (backward) scattering of $f$
with respect to the incident $e^-$ direction.

{\em {\bf Exercise:} Use (\ref{intF}-\ref{intB}) in order to
derive the second equality in (\ref{afb}) and then
check the numerical values for $A_{FB}^f$
from Table~\ref{table_zpole} for $f=\ell,b,c$. }

\subsection{Left-right asymmetry}

The left-right asymmetry is defined as
\begin{equation}
A_{LR}^f \equiv \frac{1}{P_e}
  \frac{\sigma^f(-|P_e|)-\sigma^f(+|P_e|)}
       {\sigma^f(-|P_e|)+\sigma^f(+|P_e|)} = A_e,
\end{equation}
where $\sigma^f(P_e)$ is the total (integrated over all angles)
cross-section for producing $f\bar{f}$ pairs with an electron beam
of polarisation $P_e$.

{\em {\bf Exercise:} Derive the second equality.}

\subsection{Left-right forward-backward asymmetry}

The left-right forward-backward asymmetry is defined as
\begin{equation}
\bar{A}_{FB}^f \equiv 
\frac{\sigma^f_F(-|P_e|)-\sigma_B^f(-|P_e|)-\sigma^f_F(+|P_e|)+\sigma_B^f(+|P_e|)}
     {\sigma^f_F(-|P_e|)+\sigma_B^f(-|P_e|)+\sigma^f_F(+|P_e|)+\sigma_B^f(+|P_e|)}
= \frac{3}{4}P_eA_f.
\end{equation}
It allows a direct determination of the $A_f$ quantities which are presented in 
Table~\ref{table_zpole}.

{\em {\bf Exercise:} Derive the second equality.}

To summarize or discussion so far, there are three observable
asymmetries, measuring either $A_e$, $A_f$ or their product.

\subsection{Tau polarization}

The tau lepton is the only fundamental fermion whose polarization is
experimentally accessible at LEP and SLC. The average polarization of
$\tau$ leptons is defined by
\begin{equation}
P_\tau(s) = \frac{\sigma^{tot}_+(s)-\sigma^{tot}_-(s)}{\sigma^{tot}(s)},
\end{equation}
where $\sigma^{tot}(s)$ is the $\tau$ production cross-section
{\em via} $Z$ exchange, integrated over all angles, and 
the subscripts $+$ and $-$ refer to the $\tau$ helicity states
$+1$ and $-1$. The dependence of $P_\tau$ at the $Z$ pole 
on the polar angle $\theta$ has the form~\cite{Altarelli:hv}
\begin{equation}
P_\tau(\cos\theta) = -
\frac{A_\tau(1+\cos^2\theta)+2A_e\cos\theta}{1+\cos^2\theta+2A_\tau A_e \cos\theta}.
\label{ptau}
\end{equation}

\begin{figure}[tbp]
\centerline{\epsfxsize=4.1in\epsfbox{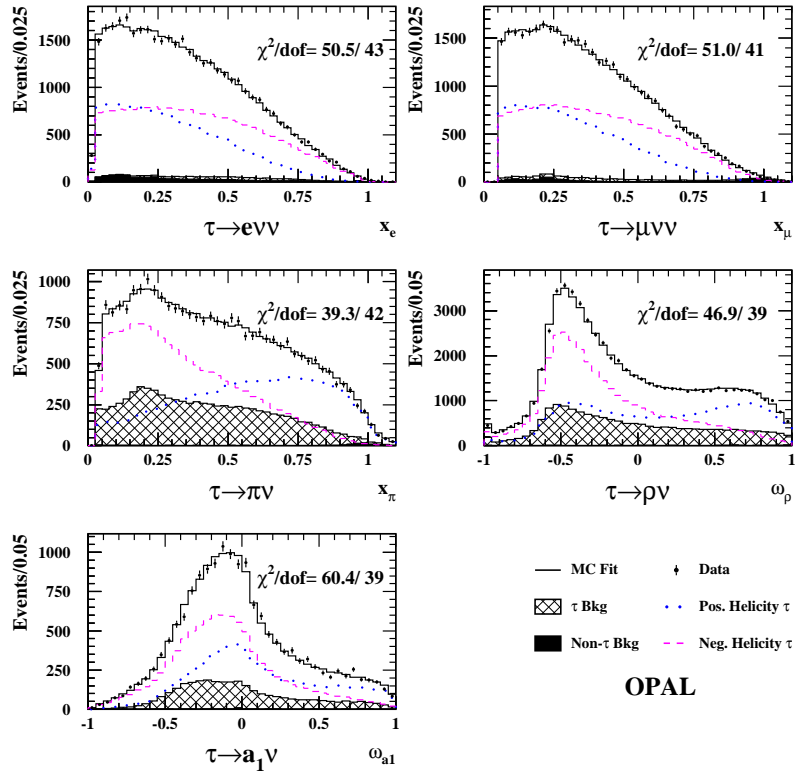}}   
\caption{Observed distributions of the polarization estimators
for the different channels studied by OPAL~\protect\cite{Abbiendi:2001km},
compared to the SM expectation for positive or negative $\tau$ helicity.}
\label{ptau_opal}
\end{figure}

The $\tau$ polarization is measured using five exclusive
decay modes which comprise about 80\% of all $\tau$ decays:
$e\nu\bar{\nu}$, $\mu\nu\bar{\nu}$, $\pi(K)\nu$,
$\rho\nu$ and $a_1\nu$. The single $\pi$ and $K$ modes are 
not normally distinguished. The different channels do not all
have the same sensitivity to the $\tau$ polarization, 
$\pi(K)\nu$ being the best in that respect (see Fig.~\ref{ptau_opal}).
The energy and/or angular distributions of the decay products
can be used as $\tau$ {\em polarization analysers}~\cite{Davier:1992nw}. 
Figure~\ref{ptau_opal} shows an example of the different distributions
used and their sensitivity to the $\tau$ polarization. 

\begin{figure}[tb]
\centerline{\epsfxsize=4.1in\epsfbox{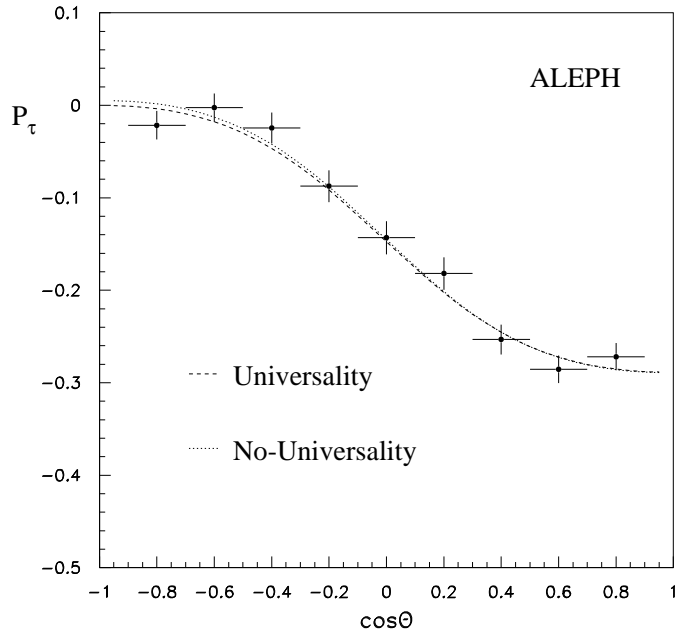}}   
\caption{Tau polarization as a function of $\cos\theta$. The dashed
(dotted) line is the result from a global fit to the data with 
(without) the universality assumption ${\mathcal A}_e={\mathcal A}_\tau$.}
\label{ptau_aleph}
\end{figure}

Fitting the experimental data for $P_\tau(\cos\theta)$ 
(an example from the ALEPH experiment~\cite{Heister:2001uh} 
is shown in Figure~\ref{ptau_aleph}) 
to Eq.~(\ref{ptau}) allows an extraction of $A_\tau$ and $A_e$
which enter Table~\ref{table_zpole} 
as $A_\tau ({\mathcal P}_\tau)$ and $A_e ({\mathcal P}_\tau)$, correspondingly.

\section{Precision Measurements at LEP-II}

LEP-II was able to pair produce $W$'s and perform measurements of 
$M_W$, $\Gamma_W$ and the $W$ branching fractions. 
There are three diagrams contributing to $W$-pair production:
a $t$-channel $\nu$ exchange, and an $s$-channel $\gamma^\ast$ and
$Z$ exchange. The latter two diagrams contain triple gauge boson couplings,
therefore LEP also tested the nonabelian nature of the SM gauge interactions.

\begin{figure}[tbp]
\centerline{\epsfxsize=4.1in\epsfbox{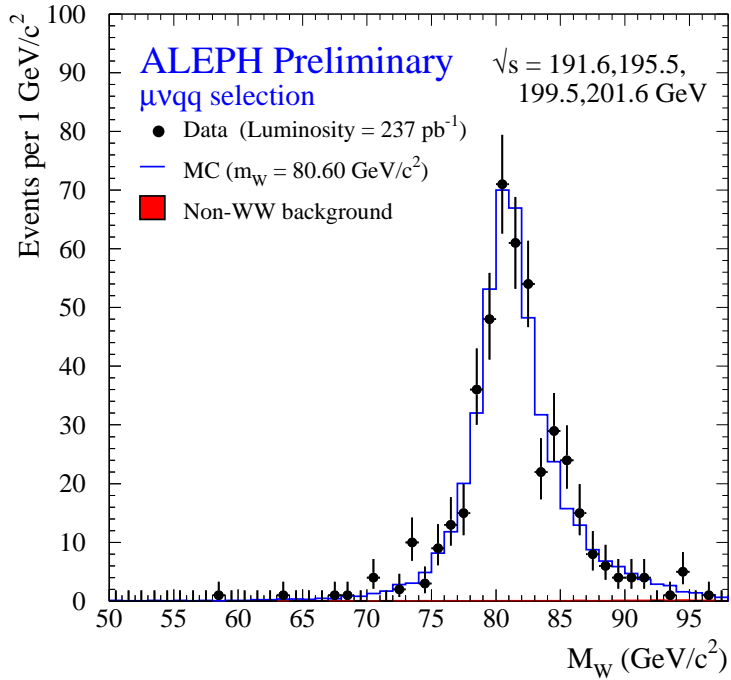}}   
\caption{Reconstructed $W$ mass distribution from the ALEPH experiment
for the $qq\mu\nu_\mu$ channel. The data are compared to the Monte Carlo 
prediction for $M_W=80.60$ GeV.}
\label{MWqqln}
\end{figure}

The $W$ mass measurement involves reconstruction of the invariant mass
of the $W$ decay products. For $WW$, there are three possible final states:
$4q$, $qq\ell\nu_\ell$ and $2\ell2\nu_\ell$. 
The LEP measurements used only the first two channels.
In the case of $qq\ell\nu_\ell$, the energy and momentum 
of the missing neutrino are easily deduced from energy 
and momentum conservation. Then the four objects are paired
up and the invariant mass of each pair is computed.
The resulting invariant mass distribution 
in the $qq\mu\nu_\mu$ channel is shown in Figure~\ref{MWqqln}.
The mass peak is very pronounced, with virtually no background.
The position of the peak allows an extraction of $M_W$, while its width
is indicative of $\Gamma_W$, where in addition one has to worry about detector 
resolution effects (smearing). All four experiments are consistent
and so far the LEP $M_W$ measurement is slightly better than the 
$M_W$ measurement from the Tevatron
(see below). The $W$ mass measurement is of extreme importance -
as we shall see later, $M_W$ is one of the observables most sensitive to the
Higgs mass $m_h$. 

The $W$ branching ratios were measured in $WW$ production and the
results displayed in Table~\ref{Wbr} nicely
demonstrate lepton universality.
\begin{table}[tbh]
\tbl{LEP measurements of the $W$ branching fractions derived from 
$WW$ production cross-section measurements\protect\cite{unknown:2003ih}. 
\label{Wbr}}
{
\begin{tabular}{|c|c|}
\hline 
Decay mode & Branching fraction (\%) \\ 
\hline
$W\rightarrow e\nu_e$        & $10.59\pm0.17$ \\
$W\rightarrow \mu\nu_\mu$    & $10.55\pm0.16$ \\
$W\rightarrow \tau\nu_\tau$  & $11.20\pm0.22$ \\
$W\rightarrow qq'$           & $67.77\pm0.28$ \\
\hline
\end{tabular}
}
\end{table}

\begin{figure}[tbp]
\centerline{\epsfxsize=4.1in\epsfbox{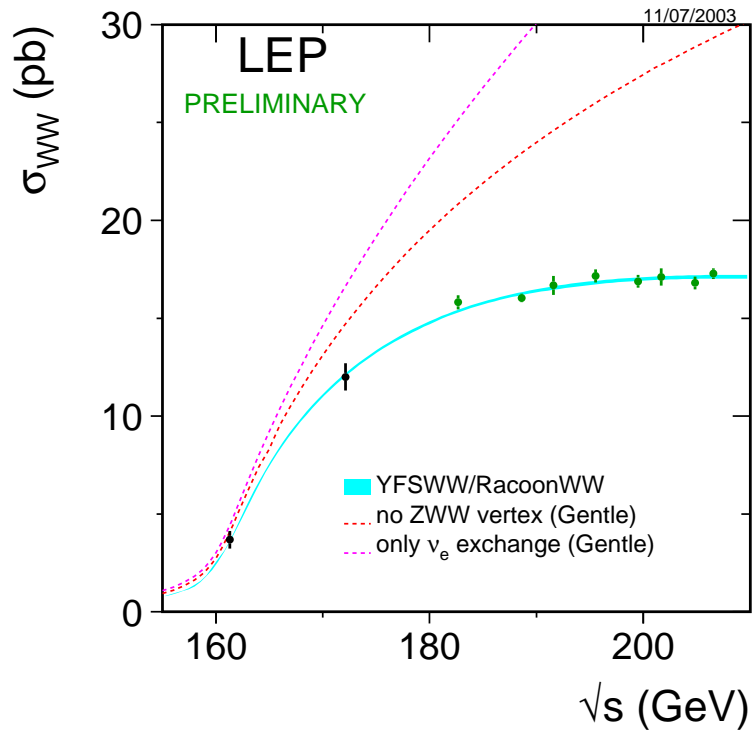}}   
\caption{The measured $WW$ cross-section at LEP as a function
of the center-of-mass energy, compared to the predictions of the
event generators RacoonWW and YFSWW. The magenta (upper dotted) line
is the theoretical result ignoring both $WW\gamma$ and $WWZ$ couplings,
while the red (lower dotted) line neglects $WWZ$ only.}
\label{wwxsec}
\end{figure}

The measurement of the rise in the $W$-pair cross-section near threshold 
provides an alternative method for determining $M_W$.
The result for energies up to 207 GeV is shown in Figure~\ref{wwxsec}.
The method is statistically limited and yields
inferior precision compared to direct $M_W$ reconstruction. 
However, it clearly demonstrates the need for 
all triple gauge boson couplings in order to understand the data.

\begin{figure}[tbp]
\centerline{\epsfxsize=4.1in\epsfbox{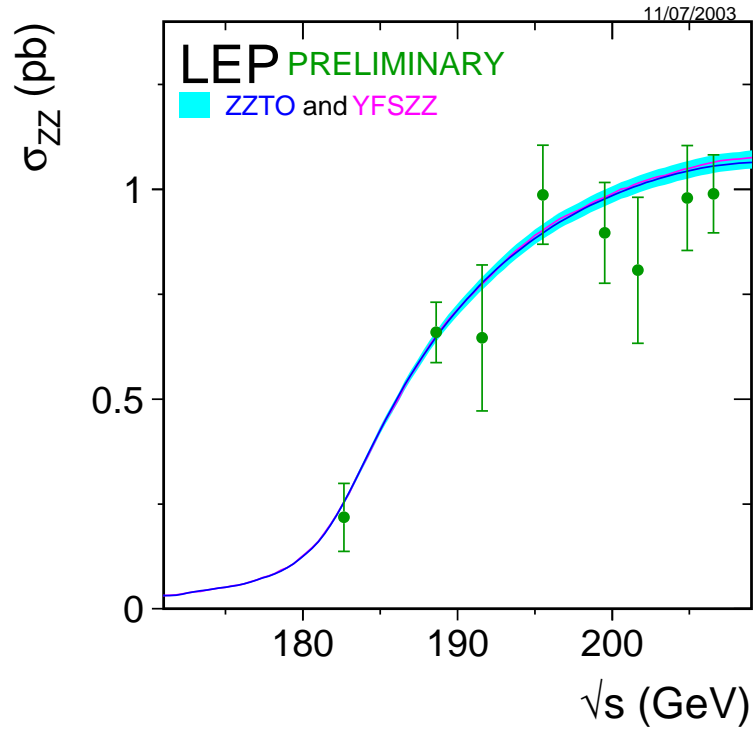}}   
\caption{The measured $e^+e^-\to ZZ$ cross-section at LEP as a function
of the center-of-mass energy. The solid line is the SM prediction
and the band indicates its $\pm2\%$ uncertainty.}
\label{zzxsec}
\end{figure}

LEP-II also measured the $ZZ$ cross-section. However, the electron couplings 
to the $Z$ are smaller, and in addition $M_Z>M_W$. Thus the sample
of $ZZ$ pairs was much smaller and the measurement of the $ZZ$ cross-section
was not as good as for the $WW$ case. Nevertheless, a comparison of the
combined data to theory (Fig.~\ref{zzxsec}) reveals good agreement 
within the large errors of the data.

\section{Precision Measurements at the Tevatron}

\subsection{Top mass measurement}

Since the top quark is so heavy, so far it can only be produced at the Tevatron.
The dominant mechanism for pair-production is $q\bar{q}\rightarrow t\bar{t}$
through a virtual $s$-channel gluon. Single top production is also possible.

The top quark decays as $t\rightarrow Wb$ almost 100\% of the time.
Top quark pair-production then gives $W^+W^-b\bar{b}$ events
and there are three possible discovery channels, depending on the $W$ decays:
\bit
\item Both $W$'s decaying hadronically - 6 jet final state ($2b4j$). This is
the largest signal cross-section, but with large QCD backgrounds.
\item One $W$ decays hadronically, the other leptonically: $\ell 2b2j\met$
(lepton plus jet sample). This was the best channel for Run I, because there was
a sufficient number of events, yet the background was under control.
\item Both $W$'s decaying leptonically: $2\ell 2b \met$ (dilepton sample).
This channel has the largest $S/\sqrt{B}$ ratio, however 
it is limited by statistics.
\eit

\begin{figure}[tbhp]
\centerline{\epsfxsize=3.9in\epsfbox{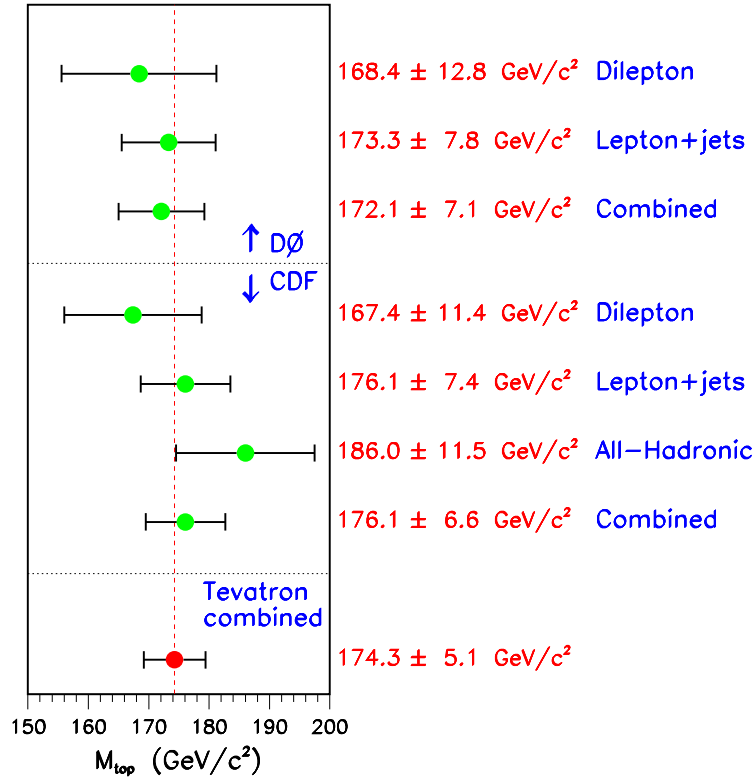}}   
\caption{Run I Tevatron results for $m_t$ and the global average.
(From Ref.~\protect\refcite{Chakraborty:2003iw}.)}
\label{mt}
\end{figure}

Both experiments (CDF and D0) have conducted measurements of $m_t$ in Run I 
and the results from the two experiments as well as from different channels
are consistent with each other (see Fig.~\ref{mt}).
The direct $m_t$ measurements from the Teavtron are in very good agreement with
the best fit value for $m_t$ extracted from the electroweak precision fits 
(see Section~\ref{sec:mw}). Right now $m_t$ is the best measured of all quark masses.

In Run II one can anticipate significant improvements -- 
the expected precision on the $m_t$ measurements is 
shown in Table~\ref{table_mt}.
In fact, there are already some preliminary top results 
from Run II \cite{Wagner:2003cv}.

\begin{table}[tbh]
\tbl{Expected precision (in GeV) on the $m_t$ measurements in Run II
for the lepton plus jet and for the dilepton channel.
(From Ref.~\protect\refcite{Baur:2002gp}.)
\label{table_mt} }
{
\begin{tabular}{|c|c|c|c|c|}
\hline 
\multicolumn{2}{|c|}{$\int{\mathcal L}dt$ (${\rm fb}^{-1}$)} & 2  & 15 & 30  \\ 
\hline
                &  Statistical   & 1.7  & 0.63 &  0.44 \\  \cline{2-5}
$bb\ell jj\met$ &  Systematic    & 2.1  & 1.2  &  1.1  \\  \cline{2-5}
                &  Total         & 2.7  & 1.3  &  1.2  \\  \cline{2-5}
\hline
                &  Statistical   & 2.4  & 0.87 &  0.62 \\  \cline{2-5}
$bb2\ell \met$  &  Systematic    & 1.4  & 1.0  &  1.0  \\  \cline{2-5}
                &  Total         & 2.8  & 1.3  &  1.2  \\  \cline{2-5}
\hline
\end{tabular}
}
\end{table}

\subsection{$W$ mass measurement}

$W$ bosons can be singly produced at the Tevatron with a relatively 
large cross-section $\sim 1$ nb. The $W$ is subsequently detected 
through its leptonic decay mode to a charged lepton and the 
corresponding neutrino. (The hadronic decays cannot be
used because of the enormous dijet background from QCD.)
The signature is $\ell\met+X$. Unfortunately, we cannot reconstruct the
invariant mass of the $W$, because of the unknown longitudinal component of the
neutrino momentum. Hence we must extract $M_W$ from transverse quantities only.

There are several possible methods~\cite{unknown:2003sv}: 
looking at the transverse mass ($M_T^W$),
the transverse momentum of the charged lepton ($p_T^\ell$) or the missing
transverse energy $\met$. Since different methods have different systematic
errors, it is desirable to have as many independent measurements as possible.

\begin{figure}[tbhp]
\centerline{\epsfxsize=4.1in\epsfbox{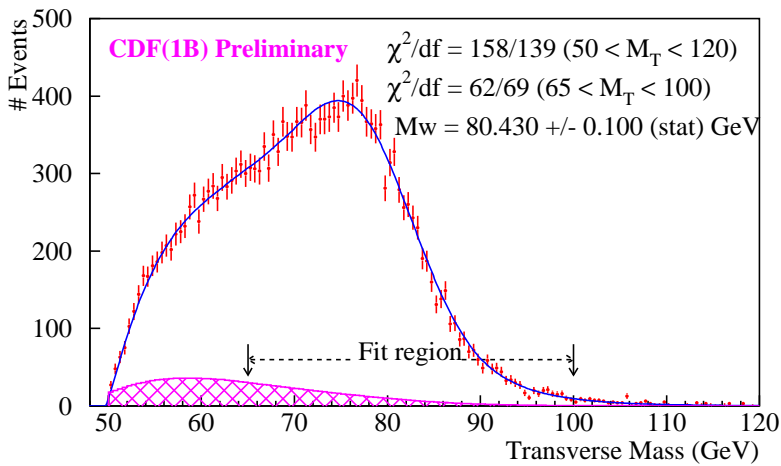}}   
\caption{Transverse mass distribution of $W\to \mu\nu$ events from the
CDF Run IB data, with the best fit. The shaded distribution is the background.}
\label{mw_t}
\end{figure}

\begin{figure}[tbhp]
\centerline{\epsfxsize=4.1in\epsfbox{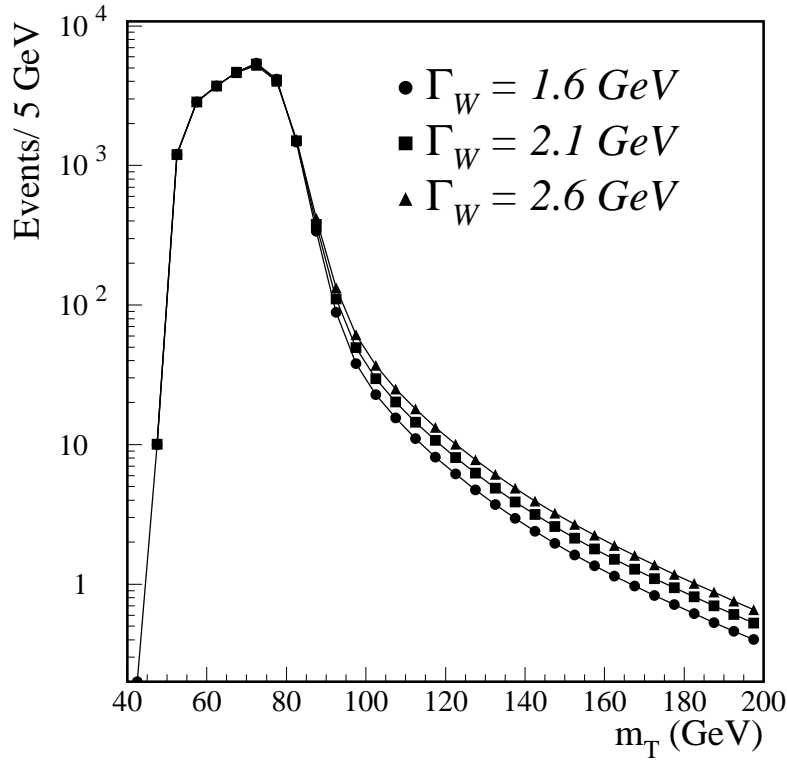}}   
\caption{Monte Carlo simulations of the transverse mass spectrum for 
different $W$-boson widths. The normalization is arbitrary.
(From Ref.~\protect\refcite{Abazov:2002xj}.)}
\label{Gamma_W}
\end{figure}

In Run I the transverse mass method was used. (The $p_T^\ell$ method 
was limited by the number of leptonic $Z$ events -- see below.)
The transverse mass\footnote{The name is appropriate since when the $W$ has
zero longitudinal momentum $p_z^W$, the transverse mass $M_T^W$ coincides with the
invariant mass $M_W$. If, however, $|p_z^W|>0$, then $M_T^W<M_W$.} is defined as
\begin{equation}
M_T^W = \sqrt{2p_T^\ell p_T^\nu (1-\cos\phi)}
\end{equation}
where $\phi$ is the angle between $\vec{p}_T^\ell$ and 
$\vec{p}_T^\nu$ in the transverse plane. $\vec{p}_T^\nu$
is nothing but the missing transverse energy and is measured as
\begin{equation}
\vec{p}_T^\nu = - (\vec{U}+\vec{p}_T^\ell)
\end{equation}
where $U$ is total transverse momentum of the remaining hadronic activity
in the event (typically a jet from initial state radiation). 
The transverse mass distribution exhibits a characteristic drop-off
around $M_W$, as shown in Fig.~\ref{mw_t}. 
The $M_W$ measurement is done by fitting Monte Carlo predictions 
for different values of $M_W$ to the observed $M_T^W$ distribution. 
In addition, at large $M_T^W$ the shape of the distribution
is sensitive to the intrinsic $W$ width $\Gamma_W$, 
as illustrated in Fig.~\ref{Gamma_W},
which allows for a direct measurement of $\Gamma_W$:
\begin{equation}
\Gamma_W = 2.055 \pm 0.125\ GeV.
\end{equation}

An alternative method for measuring $M_W$
employs the lepton $p_T$ distribution, which
cuts off around $M_W/2$. However, unlike $M_T^W$, the 
$p_T^\ell$ distribution depends on the $W$ boost, and therefore 
on the transverse momentum $p_T^W$ with which the $W$ was produced. 
It is difficult to compute $p_T^W$ by theoretical means, especially in the
low $p_T$ region, and it is best to extract it from data. For this purpose 
the Tevatron experiments have used the observed $p_T^Z$ distribution,
which has a similar shape, but fewer events. 
Hence, in Run I this method was statistically limited,
but offers good prospects for Run II.
The expected precision of the $M_W$ measurements at the Tevatron 
by the two methods is shown in Table~\ref{table_mw}.
\begin{table}[tbh]
\tbl{Expected precision (in MeV) on the Tevatron $M_W$ measurements in Run II.
(From Ref.~\protect\refcite{Baur:2002gp}.)
\label{table_mw} }
{
\begin{tabular}{|c|c|c|c|c|}
\hline 
\multicolumn{2}{|c|}{$\int{\mathcal L}dt$ (${\rm fb}^{-1}$)} & 2  & 15 & 30  \\ 
\hline
        &  Statistical   &  19  &   7 &   5 \\  \cline{2-5}
$M_T^W$ &  Systematic    &  19  &  16 &  15 \\  \cline{2-5}
        &  Total         &  27  &  17 &  16 \\  \cline{2-5}
\hline
           &  Statistical   &  44  &  16 &  11 \\  \cline{2-5}
$p_T^\ell$ &  Systematic    &  10  &   4 &   3 \\  \cline{2-5}
           &  Total         &  44  &  16 &  12 \\  \cline{2-5}
\hline
\end{tabular}
}
\end{table}

\section{Theoretical Interpretation of the Precision Electroweak Data}

We can make use of the precision electroweak data in three differrent ways:
\bit
\item Test the Standard Model. 
\item Predict the preferred mass range of yet undiscovered particles:
top quark in the past, nowadays the Higgs boson. 
\item Point towards specific new physics models (in case of some discrepancy)
or constrain new physics models (if agreement is found).
\eit
We shall discuss each one in turn.

\subsection{Testing the Standard Model}

Having made a variety of measurements for different observables, 
we can test the SM by comparing theory to experiment. For this purpose
we need to compute the theoretical prediction within the SM for each observable.
How do we do it?

Recall that in the SM we start with the input parameters
\begin{equation}
\left\{p\right\}\equiv\left\{ \alpha, G_F, M_Z, g_3, m_h, m_t, m_b, m_c, m_s, ... \right\},
\label{SMparameters}
\end{equation}
where the first three are the parameters of the electroweak sector.
Any other electroweak quantity, e.g. $\sin^2\theta_W$, $M_W$ etc. 
can be expressed in terms of these and thus is predicted by the model. 
At tree level the expressions are simple
and only involve the electroweak inputs $\alpha, G_F, M_Z$, e.g.
the Weinberg angle can be found from
\begin{equation}
\sin2\theta_W = \left(\frac{2\sqrt{2}\pi\alpha}{G_F M_Z^2}\right)^{1/2}.
\label{sin2theta}
\end{equation}
Similarly, the $W$ mass is 
\begin{equation}
M_W = M_Z\cos\theta_W
\end{equation}
with $\theta_W$ given by (\ref{sin2theta}).

The radiative corrections, however, modify the tree level relations and
introduce the remaining (non-electroweak) parameters into the game, so that
each electroweak observable depends in principle on the full set of
input parameters (\ref{SMparameters}):
\begin{equation}
{\mathcal O}_i^{theory}(\{p\}) = {\mathcal O}_i^{tree}(\alpha,G_F,M_Z)
\left[1+\Delta_{i}(\{p\}) \right].
\end{equation}
By measuring enough electroweak observables
(with sufficient precision so that we are sensitive to the radiative corrections)
we can extract information about the values of the other parameters
outside the electroweak sector. We can then test the model for consistency by comparing
the values deduced indirectly with direct measurements of those parameters.

So the strategy is as follows. First compute the corrections in a
certain renormalization scheme, treating some parameters as fixed 
inputs (usually the best measured ones). Then perform a global fit 
to the electroweak data. This will result in ``best fit'' values
for the remaining (floating) input parameters. Then 
\begin{enumerate}
\item Compare the ``best fit'' values of the floating parameters to their 
direct measurements (if available). For example, the fit will choose
``best'' values for $M_W$, $m_t$, $\alpha_s$, which have been measured by other means 
directly. On the other hand, the best fit value for $m_h$ cannot be checked 
against experiment yet, but is perhaps the most valuable piece of information
from the global fit.
\item For the best fit values found above, compute and quote the theoretical
prediction of the SM for each observable. This is the number quoted under 
``Standard Model''
in Tables~\ref{table_zpole} and \ref{table_nonzpole}. Then compare the 
experimental and theoretical values of the observables themselves.
A large discrepancy in a certain place may signal new physics which affects
that particular measurement but not the others...
\end{enumerate}

There are two popular programs on the market which would accomplish this
program:
\begin{itemize} 
\item ZFITTER~\cite{Bardin:1999yd}, which uses the on-shell scheme;
\item GAPP~\cite{Erler:1999ug} (global analysis of particle properties),
which utilizes the $\overline{MS}$ scheme and is used for the PDG review.
\end{itemize}

\subsection{Fixed parameters}
\label{sec:fixed}

The parameters which are held fixed in the global fits are the following.

{\bf Fine structure constant $\alpha$}. It is measured at low energies, so in order to
obtain $g'(M_Z)$ and $g(M_Z)$ we need to evolve it up to the $Z$ scale:
\begin{equation}
\alpha_e(M_Z) = \frac{\alpha}{1-\Delta\alpha(M_Z)}.
\end{equation}
$\Delta\alpha$ has QED and (two-loop) QCD contributions. The latter
are denoted as $\Delta\alpha_{had}^{(5)}$ (for five active quark flavors below $M_Z$).
The global fit produces a ``best fit'' value for it~\cite{Langacker:2001ij}
\begin{equation}
\Delta\alpha_{had}^{(5)}(M_Z)=0.02778\pm0.00020,
\end{equation}
which can again be compared to theoretical estimates.
The uncertainty arises from higher order perturbative and nonperturbative corrections,
from the uncertainty in the light quark masses (most notably $m_c$) and 
from insufficient $e^+e^-$ data below 1.8 GeV.
Sometimes, however, the theoretical estimates are used
as a constraint in the fit and predetermine the best fit value.

Note that $\alpha^{-1}=137.03599976\pm0.00000050$ is much better known than 
$\alpha_e^{-1}(M_Z)=127.922\pm0.020$, which explains why it is preferred 
as an input parameter.

{\bf Fermi constant} $G_F=1.16637(1)\ 10^{-5}\ {\rm GeV}^{-2}$.
It is determined from the muon lifetime formula
\begin{equation}
\tau^{-1}_\mu = \frac{G_F^2 m^5_\mu}{192 \pi^3}
F(\frac{m_e^2}{m_\mu^2})
\left( 1+\frac{3}{5}\frac{m_\mu^2}{M_W^2}\right)
\left[1+\left(\frac{25}{8}-\frac{\pi^2}{2}\right)\frac{\alpha(m_\mu)}{\pi}
+C_2\frac{\alpha^2(m_\mu)}{\pi^2}\right],
\end{equation}
where
\begin{equation}
F(x)=1-8x+8x^3-x^4-12x^2\ln x
\end{equation}
and $C_2$ is a known number. The remaining uncertainty in $G_F$ is almost entirely
from the experimental input.

{\bf Fermion masses.} For simplicity, the fermion masses are held fixed,
with the exception of $m_t$ and $m_c$.

\subsection{Floating parameters}
\label{sec:floating}

\begin{figure}[tbp]
\centering
\centerline{\epsfxsize=4.1in\epsfbox{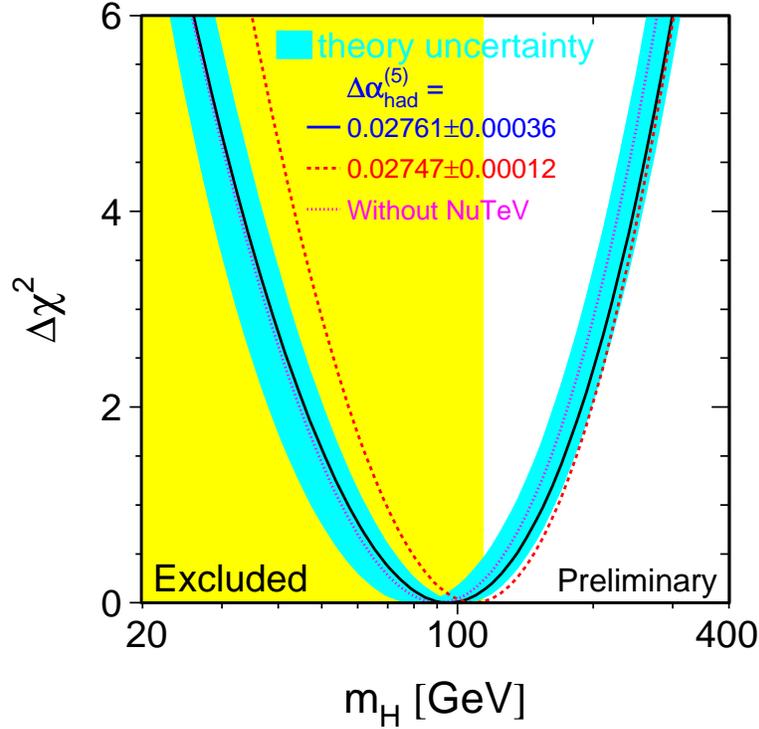}}   
\caption{Of particular interest is the constraint on the mass
$m_h$ of the Higgs boson, because this fundamental ingredient 
of the Standard Model has not been observed yet. The figure
shows the $\Delta\chi^{2}=\chi^2-\chi^2_{min}$ {\it vs.} 
$m_h$ curve derived from precision electroweak measurements,
assuming the Standard Model to be the correct theory of nature. 
The solid line is the result of the fit using all data while
the band represents an estimate of the theoretical error due
to missing higher order corrections.  The vertical band shows 
the 95\% CL exclusion limit on $m_h$ from the direct 
Higgs search at LEP. (From Ref.~\protect\refcite{unknown:2003ih}.)
}
\label{fig_blueband}
\end{figure}

The remaining parameters from the set (\ref{SMparameters}),
which were not mentioned in Sec.~\ref{sec:fixed},
are floating and their values are determined from the global fit.
\begin{itemize}
\item $M_Z$. It is measured very well, and since the
experimental measurement is included in the fit,
the ``best fit'' value for $M_Z$ is very close to the
experimental one (see Table~\ref{table_zpole}). 
\item $\alpha_s$. Its best fit value~\cite{Langacker:2003tv}
\begin{equation}
\alpha_s(M_Z)=0.1210\pm0.0018
\end{equation} 
is in very good agreement with direct determinations from tau decays,
charmonium and upsilon spectra, jet properties etc.
\item $m_t$. The best fit value, including the Tevatron data, 
is~\cite{Langacker:2003tv}
\begin{equation}
m_t = 174.2\pm4.4\ {\rm GeV}.
\end{equation} 
However, even if we take the Tevatron data out, the prediction
of precision data alone~\cite{Langacker:2003tv}
\begin{equation}
m_t = 174.0^{+9.9}_{-7.4}\ {\rm GeV}.
\end{equation} 
is in impressive agreement with the direct Tevatron value
$174.3\pm5.1$ GeV.
\item Higgs boson mass $m_h$. The best fit value is~\cite{Langacker:2003tv}
\begin{equation}
m_h = 86^{+49}_{-32}\ {\rm GeV}.
\end{equation}  
The central value is well below the LEP exclusion limit from direct searches,
but is consistent at $1\sigma$. The result for the Higgs mass is often shown as
``the blue band plot'' (Fig.~\ref{fig_blueband}). 
However, there are two caveats
which we shall discuss in more detail below. First,
the blue band plot hides a potential discrepancy between
individual contradictory measurements.
Second, the dependence on $m_h$ is only logarithmic, and new physics contributions
at the same level may have an impact on the Higgs mass prediction.
\end{itemize}

\subsection{Comparing the values for the electroweak observables}

Once we have the best fit values for the floating parameters,
we can predict the theoretical central values ${\mathcal O}^{th}$ for the 
remaining observables in Tables~\ref{table_zpole}
and \ref{table_nonzpole}.
We can then compare theory (${\mathcal O}^{th}$) to experiment 
(${\mathcal O}^{exp}$).  The results are usually 
presented as ``pulls'', which in Tables~\ref{table_zpole} 
and \ref{table_nonzpole} are defined as
\begin{equation}
{\rm pull} = \frac{{\mathcal O}^{th}-{\mathcal O}^{exp}}{\sigma^{th}},
\end{equation}
where $\sigma^{th}$ is the error on ${\mathcal O}^{th}$.

\begin{figure}[tbp]
\centering
\centerline{\epsfxsize=4.0in\epsfbox{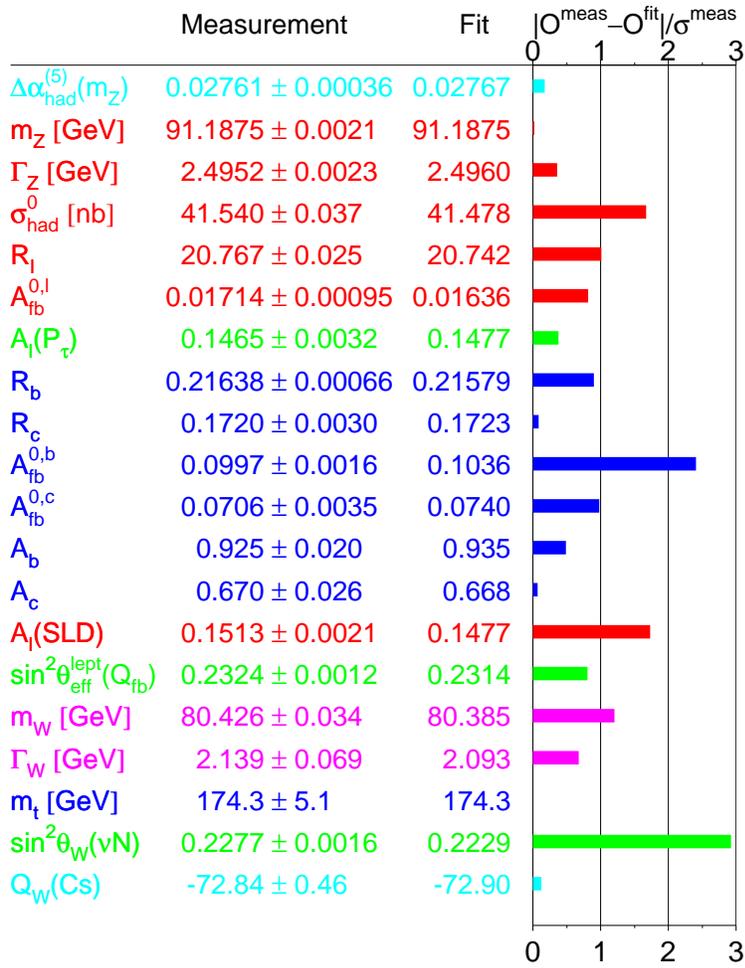}}   
\caption{Agreement between the measured values for some of the
precision electroweak observables and their 
SM predictions, as of the Summer of 2003. 
(From the LEP EWWG\protect\cite{LEPEWWG}.)}
\label{fig_pull}
\end{figure}

All in all, the fit is successful, as evidenced from Fig.~\ref{fig_pull}. 
One is tempted to conclude that the SM works pretty well, having been 
tested at the level of 1\% and less.
No major discrepancies are observed, and, 
with so many measurements, a few $2-3\sigma$ deviations 
are simply inevitable.

\subsection{The Higgs mass prediction}

\begin{figure}[tbhp]
\centering
\centerline{\epsfxsize=3.7in\epsfbox{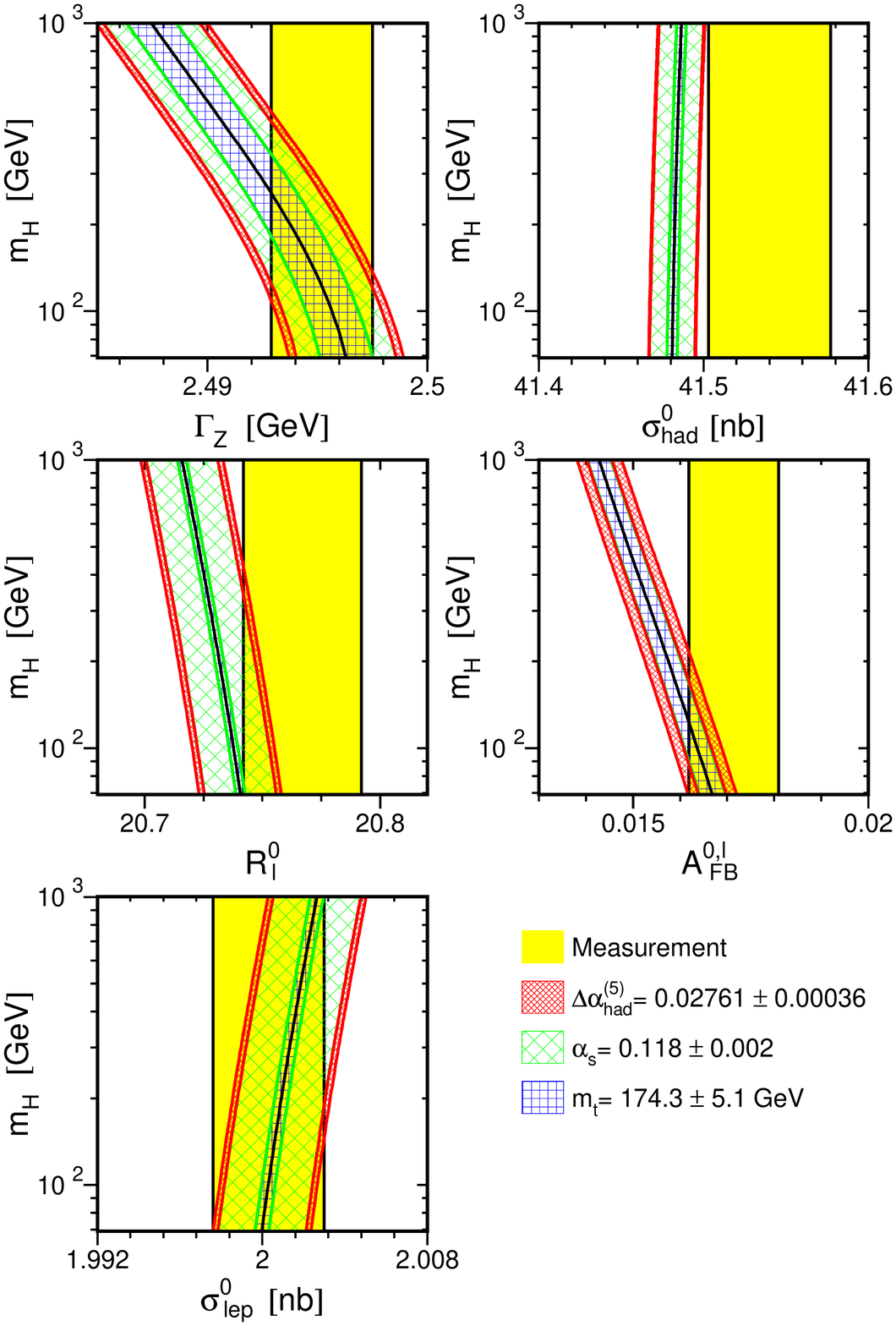}}   
\caption{Comparison of LEP-I measurements with the SM prediction 
as a function of $m_h$. The measurement with its error is shown as 
the vertical band. The width of the SM band is due to the uncertainties 
in $\Delta\alpha_{had}^{(5)}(M_Z)$, $\alpha_s(M_Z)$ and $m_t$.
The total width of the band is the linear sum of these effects.
(From the LEP EWWG~\protect\cite{unknown:2003ih}.)}
\label{higgs_sens_1}
\end{figure}

\begin{figure}[tbhp]
\centering
\centerline{\epsfxsize=3.7in\epsfbox{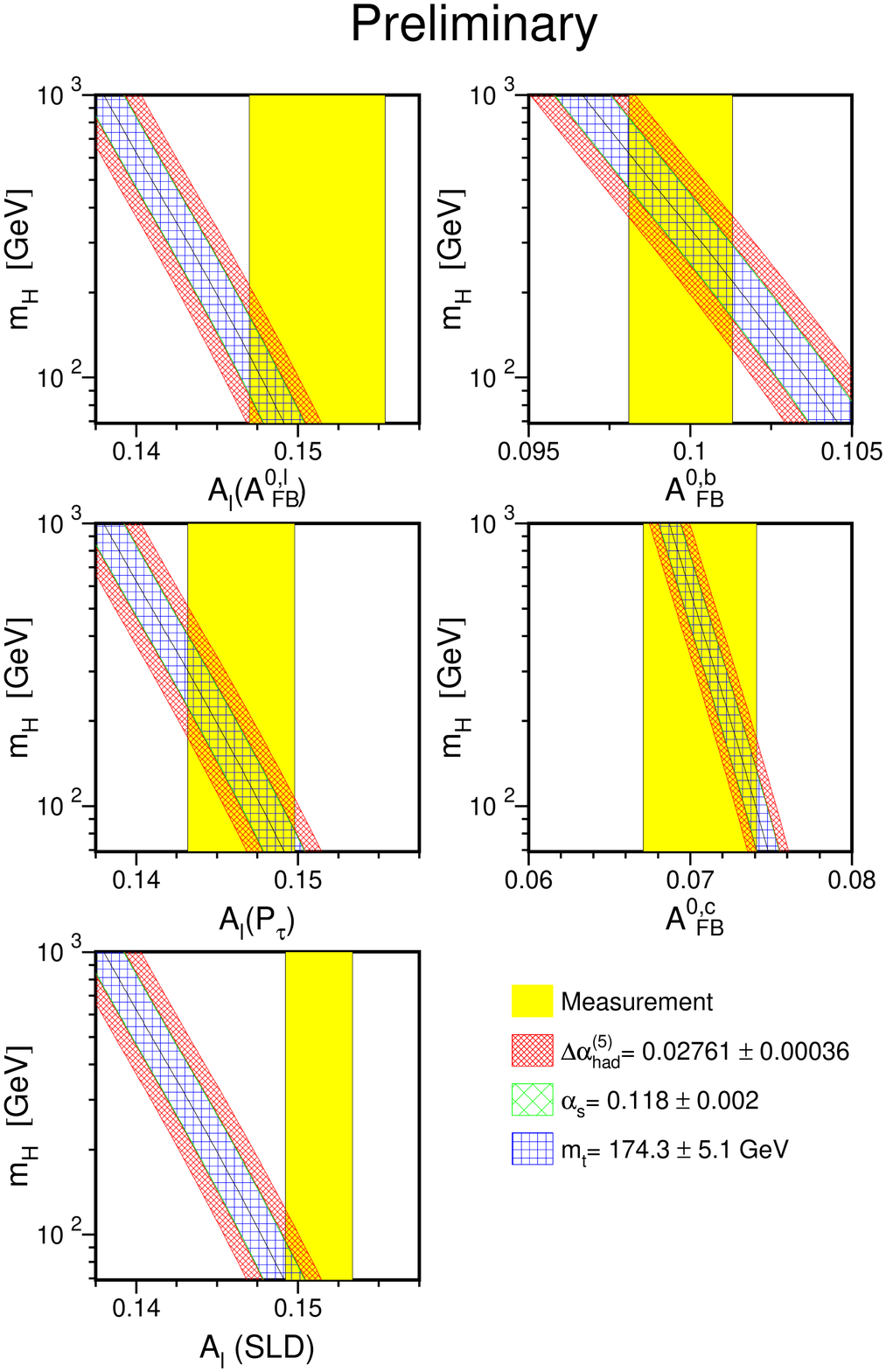}}   
\caption{Comparison of LEP-I measurements with the SM prediction 
as a function of $m_h$. The measurement with its error is shown as 
the vertical band. The width of the SM band is due to the uncertainties 
in $\Delta\alpha_{had}^{(5)}(M_Z)$, $\alpha_s(M_Z)$ and $m_t$.
The total width of the band is the linear sum of these effects.
Also shown is the comparison of the SLD measurement
of ${\mathcal A}_\ell$, dominated by $A^0_{LR}$, with the SM.
(From the LEP EWWG~\protect\cite{unknown:2003ih}.)}
\label{higgs_sens_2}
\end{figure}

\begin{figure}[tbhp]
\centering
\centerline{\epsfxsize=3.7in\epsfbox{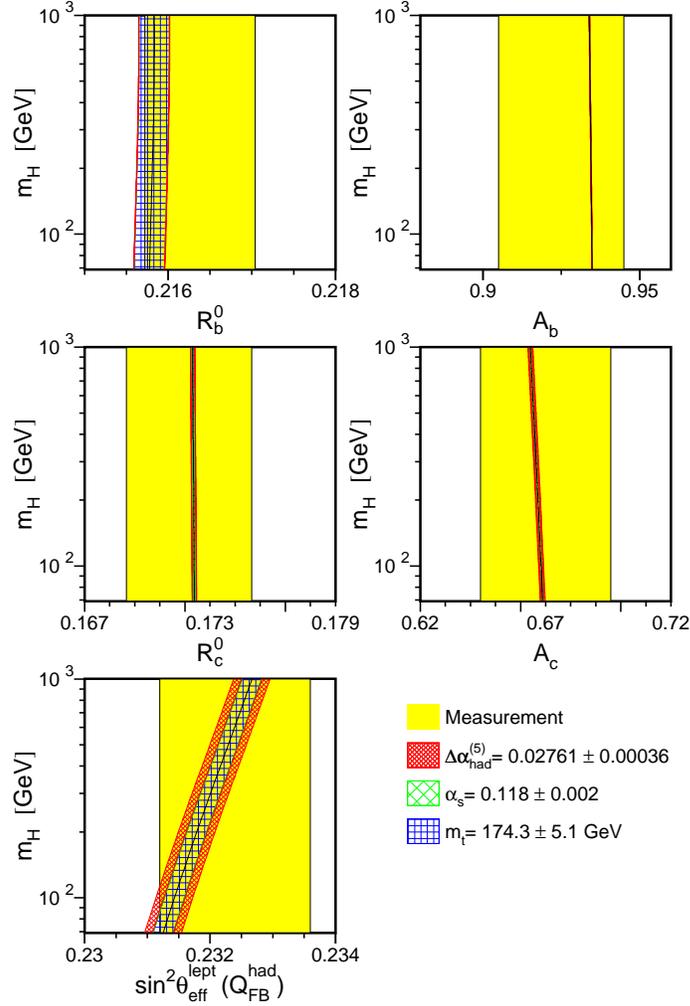}}   
\caption{Comparison of LEP-I and SLD heavy flavor measurements 
with the SM prediction as a function of $m_h$. The measurement
with its error is shown as the vertical band. The width of the
SM band is due to the uncertainties in 
$\Delta\alpha_{had}^{(5)}(M_Z)$, $\alpha_s(M_Z)$ and $m_t$.
The total width of the band is the linear sum of these effects.
Also shown is the comparison of the LEP-I measurement of the
inclusive hadronic charge asymmetry $Q^{had}_{FB}$ with the SM.
(From the LEP EWWG~\protect\cite{unknown:2003ih}.)}
\label{higgs_sens_3}
\end{figure}

\begin{figure}[tbhp]
\centering
\centerline{\epsfxsize=3.8in\epsfbox{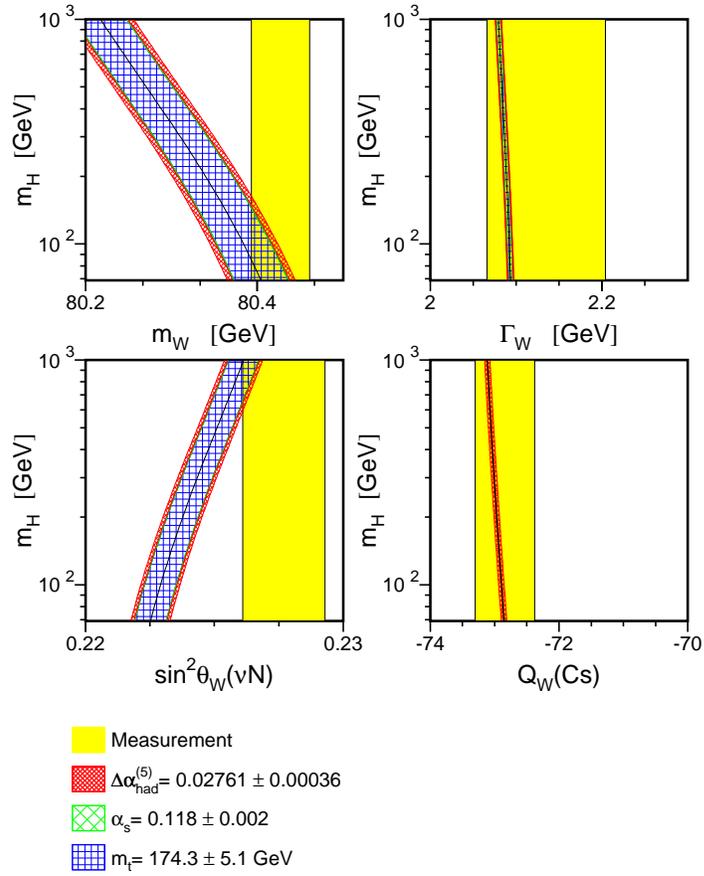}}   
\caption{Comparison of $M_W$ and $\Gamma_W$ measured at LEP-II
and $p\bar{p}$ colliders, of $\sin^2\theta_W$ measured by 
NuTeV and of atomic parity violation in caesium with
the SM prediction as a function of $m_h$. The measurement
with its error is shown as the vertical band. The width of the
SM band is due to the uncertainties in 
$\Delta\alpha_{had}^{(5)}(M_Z)$, $\alpha_s(M_Z)$ and $m_t$.
The total width of the band is the linear sum of these effects.
(From the LEP EWWG~\protect\cite{unknown:2003ih}.)}
\label{higgs_sens_4}
\end{figure}

Let us now understand better where the Higgs mass prediction comes from.
Ideally, we should concentrate on those observables which exhibit 
the strongest dependence on the Higgs boson mass $m_h$.
For this purpose, it is useful to plot the theoretical prediction 
for each observable as a function of $m_h$ and contrast it with the 
experimental measurement (see Figures~\ref{higgs_sens_1}-\ref{higgs_sens_4}).
We see that the observables most sensitive to $m_h$ 
are the $W$ mass $M_W$ and the asymmetries. We will discuss these 
in more detail in the next two subsections.

\subsection{Impact of the $M_W$ measurement on $m_h$}
\label{sec:mw}

\begin{figure}[tbhp]
\begin{center}
\centerline{\epsfxsize=4.1in\epsfbox{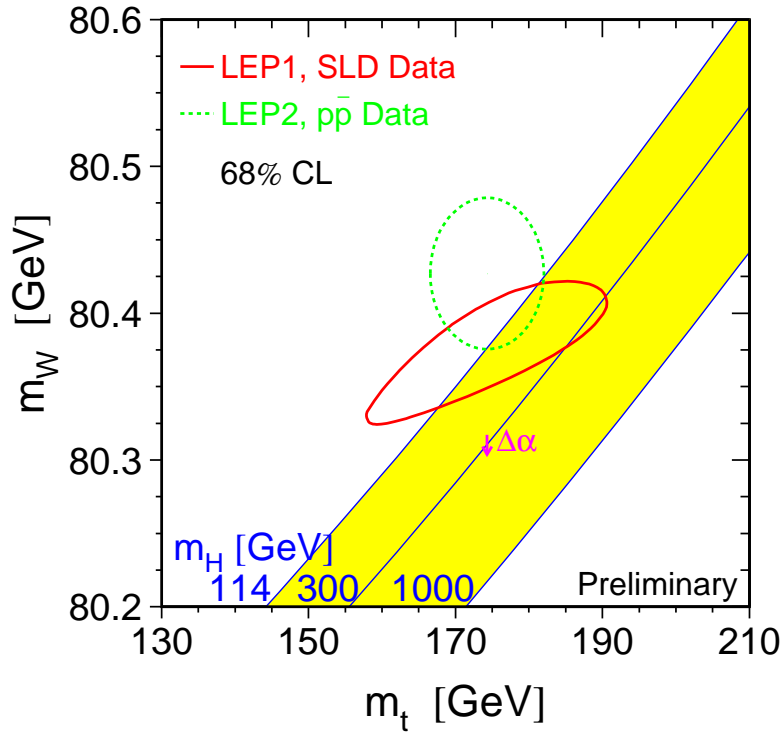}}   
\caption{A comparison of the indirect measurements of $M_W$ and $m_t$
(LEP-I+SLD data) (solid contour) and the direct measurements
($p\bar{p}$ colliders and LEP-II data) (dashed contour).
In both cases the 68\% CL contours are plotted.
Also shown is the SM relationship for the masses as a function of 
$m_h$. The arrow labelled $\Delta \alpha$ shows the variation of
this relation if $\alpha(M_Z)$ is changed by $1\sigma$.
This variation leads to an additional uncertainty to the
SM band shown in the figure. From Ref.~\protect\refcite{unknown:2003ih}. 
\label{fig_mtmw}}
\end{center}
\end{figure}

An approximate formula for $M_W$ (in the $\overline{MS}$ scheme), 
which exhibits the dependence on the relevant electroweak 
parameters, is~\cite{Sirlin:1999zc}
\begin{eqnarray}
M_W &=& 80.3827
-0.0579\ln\left(\frac{m_h}{100\ GeV}\right)
-0.008 \ln^2\left(\frac{m_h}{100\ GeV}\right) \nonumber \\
&-&0.517\left(\frac{\Delta\alpha_{had}^{(5)}(M_Z)}{0.0280}-1\right) 
+0.543\left(\left(\frac{m_t}{175\ GeV}\right)^2-1\right)\nonumber \\
&-&0.085\left(\frac{\alpha_s(M_Z)}{0.118}-1\right).
\label{sirlin}
\end{eqnarray}
The experimentally measured value for $M_W$ seems to be a bit high
(Table~\ref{table_nonzpole}). As we can see from (\ref{sirlin}),
a largish $M_W$ can be explained by either
a small $m_h$, a larger $m_t$, a smaller $\alpha_s$,
or some combination of the above.
The correlation between $M_W$, $m_t$ and $m_h$ from the
precision data is shown in  Fig.~\ref{fig_mtmw}, 
where the solid contour delineates the 68\% CL region
resulting from a global fit {\em omitting} 
the $M_W$, $\Gamma_W$ and $m_t$ measurements.
Also shown is the SM prediction for this correlation,
for $m_h=114,300,1000$ GeV. We first see that the indirect
determination of $M_W$ and $m_t$ from precision data alone
is in very good agreement with the direct $M_W$ and $m_t$ 
measurements shown with the dashed contour. 
This fact may not look that impressive, now that the top quark
and the $W$ boson have been discovered, but nevertheless it should
be considered as a triumphant success of the precision program.
We also see that both the direct and indirect measurements
of $M_W$ and $m_t$ prefer a light Higgs boson.

\subsection{Impact of the asymmetry measurements on $m_h$}

\begin{figure}[tbhp]
\centering
\centerline{\epsfxsize=4.1in\epsfbox{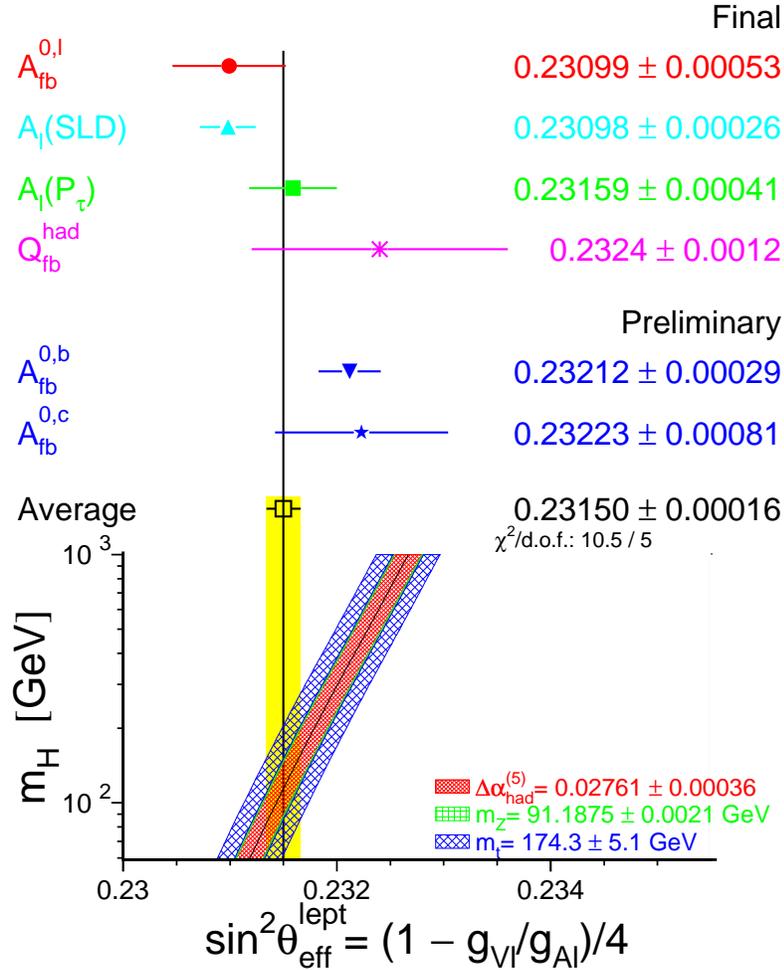}}   
\caption{Comparison of several determinations of $\sin^2\theta_{\rm eff}^{\rm lept}$ 
from asymmetries. Also shown is the prediction of the SM 
as a function of $m_h$. The width of the
SM band is due to the uncertainties in 
$\Delta\alpha_{had}^{(5)}(M_Z)$, $M_Z$ and $m_t$.
The total width of the band is the linear sum of these effects.
(From Ref.~\protect\refcite{unknown:2003ih}.)}
\label{fig_sef2}
\end{figure}

The asymmetries (or equivalently, $\sin^2\theta_{eff}$)
also exhibit significant sensitivity to $m_h$.
Examining Figure~\ref{higgs_sens_2}, we notice a couple of things. 
First, $A_\ell$ and $A_{FB}^{b,c}$ place contradictory
demands on the Higgs mass: $A_\ell$ prefers a very small $m_h$ while
$A_{FB}^{b,c}$ prefer a heavier Higgs boson. There are a couple of $A_\ell$
measurements -- one from LEP and the other from SLD, and they seem to
be in agreement. Second, since the best fit value for $m_h$ is low,
this means that $A_{FB}^{b,c}$ will be off from its 
``SM prediction''\footnote{Of course, if we compute the $A_{FB}^{b,c}$ 
prediction with a large $m_h$, then $A_{FB}^{b,c}$ will be OK, 
but a number of other well measured observables will deviate,
most notably $A_\ell$ and $M_W$, and the fit will become worse.}. 
Finally, the NuTeV measurement of $\sin^2\theta_W$ 
(Figure~\ref{higgs_sens_4}) also seems to prefer a
rather heavy Higgs boson.  The results for the various 
asymmetries are often conveniently summarized as measurements 
of $\sin^2\theta_{\rm eff}^{\rm lept}$ (Fig.~\ref{fig_sef2}).

\subsection{A Higgs puzzle?}

\begin{figure}[tbhp]
\centering
\centerline{\epsfxsize=4.0in\epsfbox{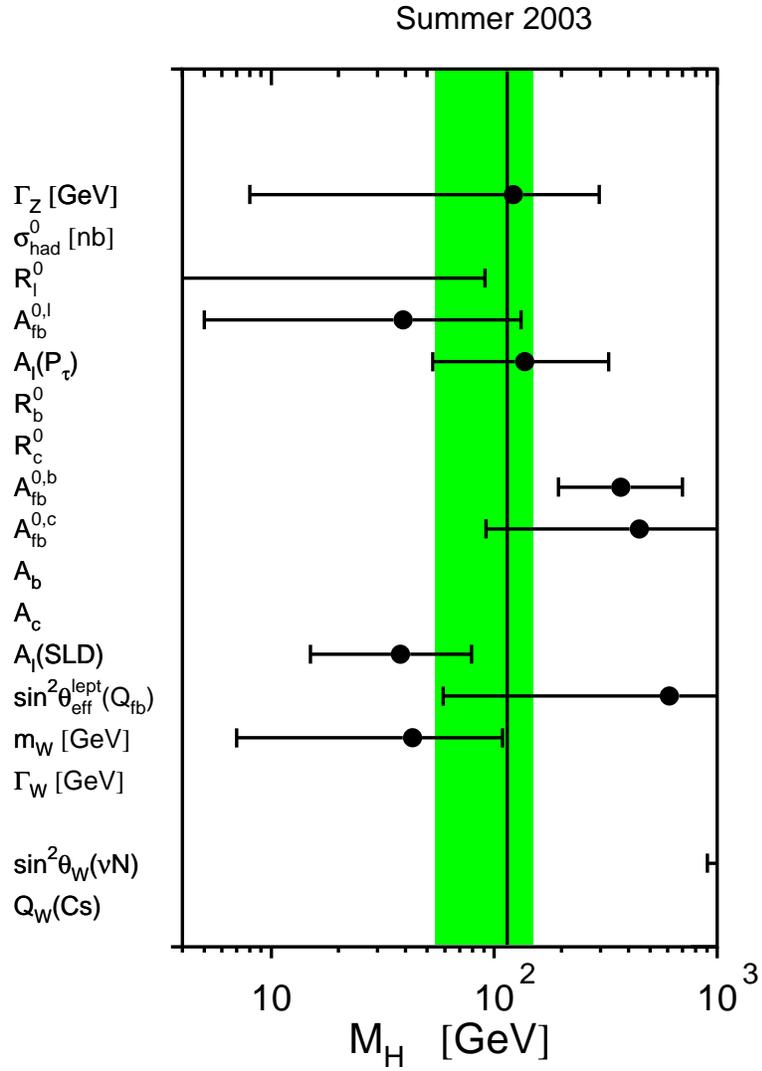}}   
\caption{Preferred range for the SM Higgs mass $m_h$ from various 
electroweak observables. The shaded band denotes the overall 
constraint on the mass of the Higgs boson derived from the full data set.
(From Ref.~\protect\refcite{unknown:2003ih}.)}
\label{fig_higgs}
\end{figure}

\begin{figure}[tbhp]
\centering
\centerline{\epsfxsize=4.1in\epsfbox{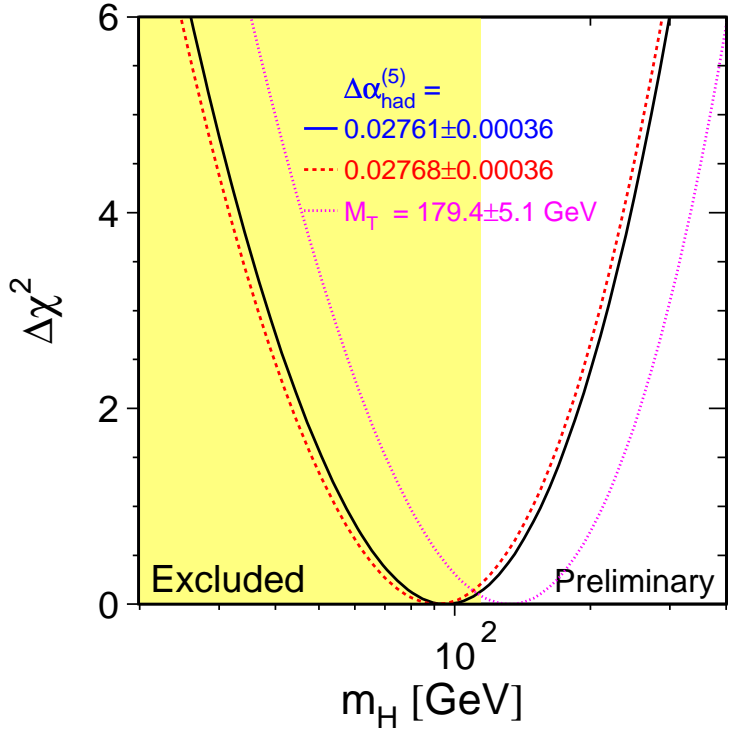}}   
\caption{The effect of a $1\sigma$ change in $m_t$ on the Higgs mass 
constraints. (From Ref.~\protect\refcite{Gambino:2003xc}.)}
\label{fig_blueband_new}
\end{figure}

Tables~\ref{table_zpole} and \ref{table_nonzpole} contain a large number
of observables, with different sensitivity to the Higgs mass $m_h$.
Let us concentrate on a few selected observables with sufficient
resolving power with respect to $m_h$ and look at the
range of $m_h$ preferred by each one (Figure~\ref{fig_higgs}). 
We see that, as discussed previously, 
$M_W$ and the leptonic asymmetries prefer a very light Higgs boson,
while $A_{FB}^{b,c}$ and NuTeV prefer a heavy Higgs boson.
The current best fit value of $m_h$ of just below the LEP limit
arises from the combination of these somewhat contradictory 
measurements. Is this a problem? Chanowitz has made the argument 
that this discrepancy presents a ``no-lose'' case for new 
physics~\cite{Chanowitz:2001bv,Chanowitz:2002cd}:
\begin{itemize}
\item One possible interpretation is that $A_{FB}^{b,c}$
is already indicative of new physics, and the
discrepancy is real. However, any such new physics effect should 
not contribute too much to $R_b$, which seems to be in 
agreement with the SM. Proposed solutions to the puzzle include
a heavy $Z'$ boson with nonuniversal couplings to the third 
family~\cite{Erler:1999nx,He:2002ha,He:2003qv} and mirror vector-like fermions
mixing with the $b_R$ quark~\cite{Choudhury:2001hs}.
\item Alternatively, the $A_{FB}^{b,c}$ (and possibly the NuTeV) measurement
could be wrong due to a statistical fluctuation or 
some unknown systematics. In that case, we should throw out 
the suspect measurements from the fit altogether. This significantly
improves the goodness of fit, but at the expense of a much lower
best fit value of $m_h$, whose upper bound from the fit 
becomes $m_h<120$ GeV at 95\% CL, 
leading to some tension with the direct LEP bound 
$m_h>114$ GeV. In this scenario new physics is again needed in 
order to modify the radiative corrections and thus 
allow an acceptable $m_h$~\cite{Altarelli:2001wx,Datta:2002jh}.
However, one should keep in mind that the tension between the
indirect determination of $m_h$ and the direct lower bound from LEP
is statistically rather weak at the moment, and may be alleviated 
significantly, if, for example, the top quark turned out to be
a little heavier, as suggested by the recent D0 analysis of 
their Run I data~\cite{Azzi:2003ct} 
(which gives $m_t=180.1\pm5.4$ GeV).
This is illustrated in Figure~\ref{fig_blueband_new},
which shows the effect of a $1\sigma$ upward fluctuation
in $m_t$ on the indirect Higgs mass determination~\cite{Gambino:2003xc}.
\end{itemize}

\section{Testing for New Physics}

The precision data can be used to constrain or point towards new physics.
To this end, there are different approaches.
\begin{itemize}
\item {\bf Model-independent.} If there is new physics, then upon integrating out the 
new heavy degrees of freedom we will end up with a set of higher dimensional operators
in the low-energy lagrangian. The most model-independent approach, therefore, will be
to write down all possible higher-dimensional operators consistent with the symmetries of the
Standard Model. These operators will be suppressed by some scale $\Lambda$ 
(presumably the scale where the new physics will appear) to the appropriate power.
The size of these operators can be parameterized by dimensionless coefficients
of order unity. We can then use the precision electroweak data in order to constrain 
the dimensionless coefficients of the relevant 
operators~\cite{Feruglio:1992wf,Appelquist:1993ka,Wudka:1994ny}. 
The advantage of this approach 
is that it is completely model-independent. Unfortunately it is too complicated and
time-consuming, and is rarely used for the case of more than a few operators.
\item {\bf Model-dependent.} Alternatively, one can completely specify the 
new physics model of interest, e.g. minimal supergravity~\cite{Erler:1998ur}, 
minimal gauge mediation~\cite{Erler:1998ur},
minimal universal extra dimensions~\cite{Appelquist:2000nn,Appelquist:2002wb}, 
little Higgs models~\cite{Csaki:2002qg,Hewett:2002px,Csaki:2003si,Gregoire:2003kr,%
Casalbuoni:2003ft,Kilic:2003mq}, etc. One can then compute the contributions
from new physics to the precision observables in terms of (\ref{SMparameters})
supplemented by the new physics model parameters $\{q\}$:
\begin{equation}
{\mathcal O}_i^{theory} = {\mathcal O}_i^{tree}(\alpha,G_F,M_Z)
\left[1+\Delta_{i}^{SM}(\{p\})+\Delta_{i}^{NP}(\{p,q\}) \right],
\end{equation}
redo the fits and derive best fit values and constraints on 
$\{q\}$ in a similar way as was done for $\{p\}$.
In full generality (i.e. full leading order corrections 
to all precision observables), the method is again very time-consuming, 
so it is usually applied for a single (or a few) observables. 
\item {\bf Oblique parameters.} A third method, which sits somewhere in between,
is the method of the so called $S,T,U$ parameters 
(also called Peskin-Takeuchi parameters~\cite{Peskin:1990zt,Peskin:1991sw}).
It amounts to making the approximation that the dominant new
physics effects reside in the gauge boson propagators (self-energies). Notice
that almost every electroweak observable involves some gauge boson propagator.
So once we compute the new physics effects on the gauge boson propagators,
we have essentially accounted for a whole class of corrections which appear
in every observable, i.e. they are universal. In this method one
neglects the process-specific corrections, i.e. the vertex and box
corrections and the fermion (and Higgs) self-energies. Apriori 
we don't know whether this approximation is justified within a specific new physics
model, however, there are many classes of models where it works. In 
scenarios with many new particles, there is a simple argument as to why
the oblique corrections are most of the story - the gauge bosons couple
to all particles charged under the corresponding gauge group, 
hence their self-energy 
corrections are enhanced by the multiplicity of the new particles.
In contrast, the flavor of the loop particles in the process-specific 
corrections is fixed by the flavor on the external legs.
To summarize, the advantages of the method are: 1) simple
calculations (there aren't too many relevant diagrams); 
2) universality -- i.e. the parameters are computed once and for all
and affect all observables; 3) if $S,T,U$ are added as free parameters to
(\ref{SMparameters}), their best fit values can be computed ahead of time 
by the fitting experts, and then supplied to model-builders, 
who in turn only need to learn to calculate $S,T,U$ within
a specific model and need not worry about the fitting procedure.
\end{itemize}

\subsection{$S$, $T$, $U$ parameters}

The $S,T,U$ parameters are defined as~\cite{Erler:ew}
\begin{eqnarray}
\widehat{\alpha}_e(M_Z) T &\equiv& 
 \frac{\Pi^{new}_{WW}(0)}{M_W^2}
-\frac{\Pi^{new}_{ZZ}(0)}{M_Z^2}, \\
\frac{\widehat{\alpha}_e(M_Z)}{4\widehat{s}^2_Z\widehat{c}^2_Z} S &\equiv& 
 \frac{\Pi^{new}_{ZZ}(M_Z^2)-\Pi^{new}_{ZZ}(0)}{M_Z^2}, \\
\frac{\widehat{\alpha}_e(M_Z)}{4\widehat{s}^2_Z} (S+U) &\equiv& 
 \frac{\Pi^{new}_{WW}(M_W^2)-\Pi^{new}_{WW}(0)}{M_W^2}.
\end{eqnarray}
where $\widehat{s}_Z^2$ is defined in terms of the $\overline{MS}$ running couplings
as $\widehat{s}_Z^2\equiv
{\widehat{g}'}{}^2(M_Z)/(\widehat{g}'{}^2(M_Z)+\widehat{g}^2(M_Z))$.
Notice that the $S,T,U$ parameters are defined so that 
for the Standard Model they are equal to 0.

\begin{figure}[tbhp]
\centerline{\epsfxsize=4.1in\epsfbox{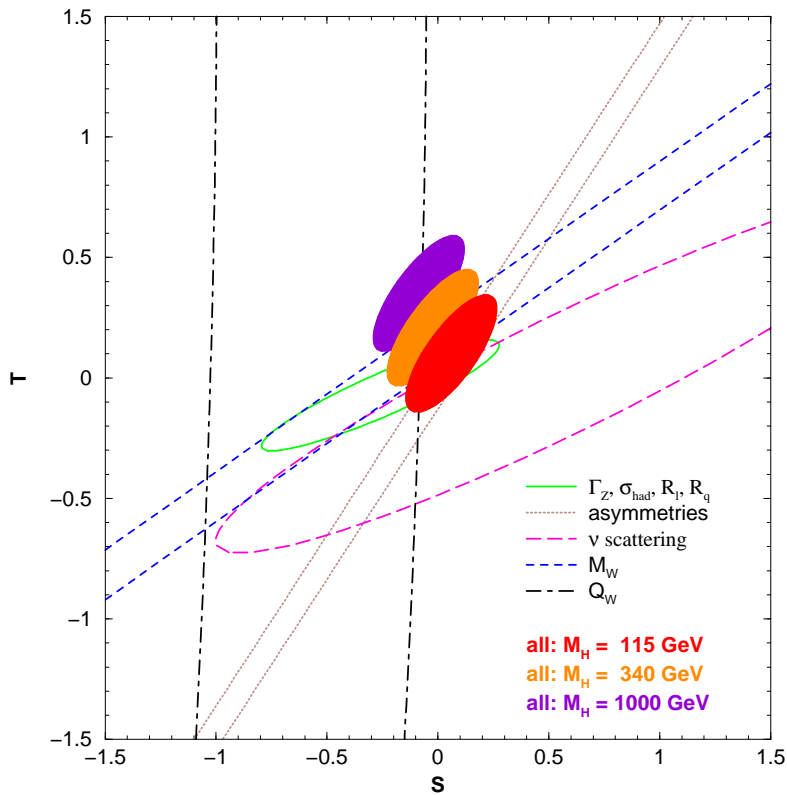}}   
\caption{$1\sigma$ constraints on $S$ and $T$ from various inputs.
The contours assume $m_h=115$ GeV except for the central and upper 
90\% CL regions (shaded, from all data) which are for $m_h=340$ GeV and
$m_h=1000$ GeV, correspondingly. All fits assume $U=0$. 
(From Ref.~\protect\refcite{Erler:ew}.)}
\label{fig_ST_langacker}
\end{figure}

One can now consider $S,T,U$ as floating parameters added to
(\ref{SMparameters}) and redo the fits to electroweak data.
The current global fit result is~\cite{Langacker:2003tv}
\begin{eqnarray}
S &=& -0.14 \pm 0.10 (-0.08),  \\
T &=& -0.15 \pm 0.12 (+0.09),  \\
U &=& +0.32 \pm 0.12 (+0.01) 
\end{eqnarray}
for $m_h = 115.6$ GeV (the parentheses show the change in going
to $m_h=300$ GeV).
Best fit contours in the $S-T$ plane (for $U=0$)
are shown in Fig.~\ref{fig_ST_langacker}.
We see that $S=T=0$ is in the region of a light Higgs boson,
i.e. in the absence of new physics, the fits prefer small $m_h$,
as we saw in the previous Section.
However, a heavy Higgs is not out of the question, provided there
are positive new physics contributions to $T$.

\subsection{Constraining new physics scenarios}

Previously we saw that precision electroweak data generally 
prefers a light SM Higgs boson (Figure~\ref{fig_blueband}),
in which case any new physics effects appearing through $S$, $T$ and 
$U$ should be small. However, this does not rule out a possible
``conspiracy''~\cite{Kolda:2000wi,Peskin:2001rw}, where the undesired
contributions from a heavy Higgs boson in the SM are fortuitously
cancelled by new physics effects. While a ``conspiracy''
is still a valid option in principle, the current thinking is that
it would involve a large degree of fine-tuning and is therefore 
theoretically disfavored. As a result, we tend to like new physics
models where no major deviations in the precision observables
are expected. A simple and effective way to guarantee this
is to have the new physics contribute only at the loop level, 
for example due to a conserved symmetry which distinguishes the
SM particles from the rest. Some well known examples are:
supersymmetry with conserved $R$-parity, 
Universal Extra Dimensions with conserved $KK$-parity~\cite{Appelquist:2000nn}, 
and little Higgs models with conserved $T$-parity~\cite{Cheng:2003ju}.
Conversely, models with {\em tree-level} contributions to
electroweak observables (e.g. generic little Higgs theories) 
tend to have trouble with the data.

\section{Concluding Remarks}

Overall, the SM is in good shape, and the agreement 
between theory and experiment has in fact improved 
since the time of TASI-2002. The results in Tables~\ref{table_zpole}
and \ref{table_nonzpole} are from 1/03 \cite{Langacker:2003tv},
while the electroweak data shown at the TASI school was from 
7/01~\cite{Langacker:2001ij}.
Some notable changes and recent updates are the following:
\begin{itemize}
\item Table~\ref{table_nonzpole} lists the Summer 2002 value
for $M_W$ ($80.447\pm0.042$ GeV), but in the Winter of 2003 
a revised ALEPH analysis lowered the LEP average to 
$80.412\pm0.042$ GeV, which is closer to the SM 
best fit prediction and will lead to a small increase 
in the best fit value for $m_h$.
\item New measurements of $M_W$ and $m_t$ are expected soon from 
Run II at the Tevatron. A new preliminary DO analysis
\cite{Azzi:2003ct} of Run I data gives $m_t=180.1\pm5.4$ GeV,
which again will raise the best fit value for $m_h$.
\item The NuTeV results on deep inelastic scattering
in terms of $\sin^2\theta_W$ ($0.2277\pm0.0016$)
are $3.0\sigma$ above the global fit value of $0.2228\pm0.0004$.
Possible explanations of the NuTeV anomaly
include an unexpectedly large violation 
of isospin in the quark sea~\cite{Davidson:2001ji,Zeller:2002du},
the effect of an asymmetric strange 
sea~\cite{Davidson:2001ji,Zeller:2002du,Bernstein:2002sa},
nuclear shadowing \cite{Miller:2002xh,Melnitchouk:2002ud} 
or NLO QCD corrections~\cite{Dobrescu:2003ta,Kretzer:2003wy}.
\item The status of the $g_\mu-2$ anomaly after the 2002
result from BNL was still somewhat uncertain, 
depending on the particular theoretical analysis used to 
extract the hadronic contribution to the Standard Model
prediction for $g_\mu-2$. In addition, last year the
CMD-2 Collaboration at Novosibirsk discovered that part of the radiative 
treatment was incorrectly applied to their $e^+e^-$ data 
(a lepton vacuum polarization diagram was omitted) 
which prompted a reanalysis of the $e^+e^-$ data~\cite{Akhmetshin:2003zn}.
This subsequently brought the theory prediction in better agreement with
experiment: the deviation was $1.9\sigma$ ($0.7\sigma$) for 
the $e^+e^-$ ($\tau$-based) estimates~\cite{Davier:2003pw,Hagiwara:2003da}.
The final result based on negative muons 
has just been announced~\cite{Bennett:2004pv}. It is consistent
with the previous measurements and brings the
world average to $a_\mu(exp)=11659208(6)\times 10^{-10}$ (0.5 ppm).
The difference between $a_\mu(exp)$ and the SM theoretical prediction
based on the $e^+e^-$ or $\tau$ data is now $2.7\sigma$ 
and $1.4\sigma$, respectively. New results from KLOE and BABAR 
may help sort out the theoretical predictions. If the discrepancy stands, 
an obvious candidate for a new physics explanation would be
supersymmetry with relatively low superpartner 
masses and large $\tan\beta$~\cite{Feng:2001tr,Everett:2001tq}.
\end{itemize}

\section*{Acknowledgments}
It is a pleasure to thank the organizers of TASI-2002
(Howard Haber and Ann Nelson) for creating a very stimulating 
atmosphere at the school, as well as the participants for their 
enthusiasm and insight. I would like to thank Andreas Birkedal 
for helpful comments on the manuscript.
This work is supported in part by the 
US DoE under grant DE-FG02-97ER41029.

%
%
%
%

\end{document}